\documentclass[fleqn,usenatbib]{mnras}
\usepackage{newtxtext,newtxmath}
\usepackage[T1]{fontenc}

\DeclareRobustCommand{\VAN}[3]{#2}
\let\VANthebibliography\thebibliography
\def\thebibliography{\DeclareRobustCommand{\VAN}[3]{##3}\VANthebibliography}

\usepackage{graphicx}	
\usepackage{amsmath}	
\usepackage{amsfonts}
\usepackage{bbold}
\usepackage{bbm}
\usepackage{subcaption}
\usepackage[dvipsnames]{xcolor}
\usepackage[export]{adjustbox}

\colorlet{linkcolor}{BrickRed}
\hypersetup{colorlinks=true,
linkcolor=linkcolor,
citecolor=linkcolor,
urlcolor=linkcolor,
,linktocpage=true
,pdfproducer=medialab}

\title[Data Compression and Inference in Cosmology with Self-Supervised Machine Learning]{Data Compression and Inference in Cosmology with Self-Supervised Machine Learning}

\author[A. Akhmetzhanova, S. Mishra-Sharma, and C. Dvorkin]{
Aizhan Akhmetzhanova,$^{1}$\thanks{E-mail: \href{mailto:aakhmetzhanova@g.harvard.edu}{aakhmetzhanova@g.harvard.edu}}
Siddharth Mishra-Sharma$^{2, 3, 1}$\thanks{E-mail: \href{mailto:smsharma@mit.edu}{smsharma@mit.edu}},
and Cora Dvorkin$^{1}$\thanks{E-mail: \href{mailto:cdvorkin@g.harvard.edu}{cdvorkin@g.harvard.edu}}
\\
$^{1}$Department of Physics, Harvard University, Cambridge, MA 02138, USA\\
$^{2}$The NSF AI Institute for Artificial Intelligence and Fundamental Interactions, Cambridge, MA 02139, USA\\
$^{3}$Center for Theoretical Physics, Massachusetts Institute of Technology, Cambridge, MA 02139, USA
}

\pubyear{2023}

\begin{document}
\label{firstpage}
\pagerange{\pageref{firstpage}--\pageref{lastpage}}
\maketitle

\begin{abstract}
The influx of massive amounts of data from current and upcoming cosmological surveys necessitates compression schemes that can efficiently summarize the data with minimal loss of information.
We introduce a method that leverages the paradigm of self-supervised machine learning in a novel manner to construct representative summaries of massive datasets using simulation-based augmentations. Deploying the method on hydrodynamical cosmological simulations, we show that it can deliver highly informative summaries, which can be used for a variety of downstream tasks, including precise and accurate parameter inference. We demonstrate how this paradigm can be used to construct summary representations that are insensitive to prescribed systematic effects, such as the influence of baryonic physics. Our results indicate that self-supervised machine learning techniques offer a promising new approach for compression of cosmological data as well its analysis.
\end{abstract}


\section{Introduction}\label{sec:Introduction}

Over the last few decades, cosmology has undergone a phenomenal transformation from a `data-starved' field of research to a precision science. 
Current and upcoming cosmological surveys such as those conducted by the \emph{Dark Energy Spectroscopic Instrument} (DESI) \citep{data_aghamousa2016desi}, \emph{Euclid} \citep{data_laureijs2011euclid}, the \emph{Vera C. Rubin Observatory} \citep{data_lsst2012large}, and the \emph{Square Kilometer Array} (SKA) \citep{data_SKA_weltman2020fundamental}, among others, will provide massive amounts of data, and making full use of these complex datasets to probe cosmology is a challenging task. The necessity to create and manipulate simulations corresponding to these observations further exacerbates this challenge.

Analyzing the raw datasets directly is a computationally expensive procedure, so they are typically first described in terms of a set of informative lower-dimensional data vectors or \textit{summary statistics}, which are then used for parameter inference and other downstream tasks.
These summary statistics are often motivated by inductive biases drawn from the physics of the problem at hand. Some widely used classes of summary statistics include power spectra and higher-order correlation functions \citep{IntroStats_Bispectra_CMBxLSS,IntroStats_PkBiTrispectra,IntroStats_BossBispectra,Chen:2021vba}, wavelet scattering transforms \citep{2020MNRAS.499.5902C,IntroStats_Wavelets,Valogiannis:2022xwu}, overdensity probability distribution functions \citep{IntroStats_PDF}, void statistics \citep{IntroStats_Voids_Pisani2019,IntroStats_Voids_Hawken2020}, and many others. While these statistics have been successful in placing tight constraints on cosmological models, the sufficiency of manually-derived statistics (i.e., ability to compress all physically-relevant information) is the exception rather than the norm.

In addition, given the estimated sizes of the datasets from future surveys, even traditional summary statistics might be too large for scalable data analysis. For instance, \cite{heavens2017_summaryStats_estimate_massive} estimates that for surveys such as {\it Euclid} or LSST, with an increased number of tomographic redshift bins, the total number of data points of summary statistics for weak-lensing data (such as shear correlation functions) could be as high as $\sim 10^4$, which might be prohibitively expensive when the covariance matrices for the data need to be evaluated numerically from complex simulations. 
In order to take advantage of the recent advances in the field of simulation-based inference (SBI) (e.g. \citet{sbi_frontierSBI_cranmer2019,pydelfi_Alsing_2019}), the size of the summary statistic presents an important consideration due to the curse of dimensionality associated with the comparison of the simulated data to observations in a high-dimensional space \citep{alsing2019nuisance}.

A number of methods have been proposed to construct optimal statistics that are compact yet retain all the relevant cosmological information. Some have focused on creating compression schemes that preserve the Fisher information content of the original dataset. 
One approach that has been widely applied in astronomy and cosmology is the Massively Optimised Parameter Estimation and Data (MOPED) compression algorithm which compresses the entire dataset to $M$ data points, where $M$ is the number of parameters of a physical theory or a model \citep{heavens2000massiveCompression}. 
\cite{Alsing17_GenMassiveOptimalDataCompression} found that, for non-Gaussian data, one can still compress the dataset down to $M$ data points optimally by using the score function. The score function, in this case defined as the derivative of the log likelihood with respect to the parameters of the model, provides optimal compressed statistics for the linearized log-likelihoods not limited to be Gaussian. 
\cite{zablocki2016extreme} and \cite{alsing2019nuisance} extended this approach further to design a compression scheme to obtain `nuisance-hardened' summaries of the original dataset from the score statistics. This further reduces the size of the summaries from $M$ points to $N$ points where $N$ is the number of parameters of interest in the model. These summaries, which can also be defined as the score function of the nuisance-marginalized log-likelihood, preserve the Fisher information corresponding to the parameters of interest and are optimal for Gaussian likelihoods. 

There have also been considerable efforts to develop and apply novel machine learning methods in order to find optimal summary statistics. \cite{IMNNs2018} introduced a method called Information Maximising Neural Networks (IMNNs), which trains a neural network to learn informative summaries from the data by using the (regularized) Fisher information as an objective function. This approach also requires a choice of fiducial parameter values, although later works \citep{IMNNs_makinen_fields,IMNNs_makinen_graph} have found that IMNNs can be robust to shifts from fiducial parameter points. Another direction explored compression schemes that optimize an estimate of the mutual information between the summaries and the parameters of interest \citep{MI_example_Jeffrey}. 
These methods can produce globally-sufficient summaries valid for a range of parameters within the of support of available simulations and not only for a set of fiducial values. 

In this work, we extend and explore self-supervised machine learning techniques as a complementary approach to derive compressed summary statistics. 
Self-supervised learning leverages the structure and internal symmetries of the data to learn informative summaries without explicit need for labels.
In astrophysics, self-supervised learning has been applied to a variety of downstream tasks, including galaxy morphology classification and photometric redshift estimation (e.g. \citet{SSL_astronomy,2020MNRAS.499.5902C,SSL_contrastive_radioGalaxy}), with extensions to domain adaptation \citep{SSL_astro_domainAdapt_Ciprijanovic,SSL_astro_domainAdapt_VegaFerrero} and neural posterior estimation \citep{SSL_astro_grav_waves}. Another set of recent works has focused on using compressed summary statistics more broadly for astrophysical data exploration, anomaly detection, and self-supervised similarity search \citep{SSL_astronomy_SelfSimilaritySearch,SSL_astro_data_exploration,SSL_astro_domainAdapt_VegaFerrero,2023arXiv230807962D}. We refer the reader to \citet{SSL_contrastive_astro_reviews} for a more extensive review of the application of contrastive self-supervised learning methods in astrophysics.
We extend the self-supervision paradigm in a novel manner, using physically-motivated simulation-based augmentations to inform the construction of summary representations.
We investigate the potential of our method for compressing cosmological datasets, such as density maps, into informative low-dimensional summaries and their downstream use for parameter inference.

This paper is organized as follows. In Sec. \ref{sec:SSL}, we review the framework of self-supervised learning, contrasting this paradigm with traditional supervised learning. We then describe the particular self-supervised method used in this study, VICReg \citep{bardes2021vicreg}. In Sec. \ref{sec:VICreg_data_compression}, we showcase the performance of our method for data compression and downstream parameter inference through case studies using mock lognormal density fields, as well as more complicated simulations -- matter density maps from the CAMELS suite \citep{CAMELS}. We compare the method's performance to an equivalent supervised baseline and, where applicable, to theoretical expectation based on Fisher information. In Sec. \ref{sec:BaryonicMarg}, we explore our method's potential to construct summaries that are insensitive to nuisance parameters and systematic effects, e.g. the influence of baryonic physics, while preserving information pertaining to relevant aspects of the model, e.g. cosmological parameters. Section \ref{sec:application_to_SBI} presents an application of our compression scheme for sequential simulation-based inference via generative emulation. Finally, we conclude in Sec. \ref{sec:Conclusions} with a summary discussion and an outlook for future work and improvements.

\section{Methods}\label{sec:SSL}

\subsection{Self-supervised learning}

Self-supervised learning has recently emerged as a powerful framework for learning meaningful representations across a wide variety of data modalities without the need for explicit labels \citep{SSL_review}. 
Self-supervised learning methods have also been shown to achieve performance comparable to fully-supervised baselines on a variety of downstream tasks such as, in the context of computer vision, image classification and object recognition (e.g. \citet{SSL_MoCo,chen2020simple}). 

The pipeline of self-supervised learning usually involves two steps. 
In the first step, often referred to as pre-training, an \textit{encoder} network is trained to learn representations, or summaries, of the data which are invariant under various transformations or augmentations.
This training step does not require the input dataset to come with labels, which is particularly advantageous in cases when obtaining or scaling up labelled datasets is expensive. 
Another important aspect of self-supervised learning is that, since training of the encoder network does not rely on labels, it can leverage the structure of the data itself to learn useful and informative summaries or representations of the data.  

After pre-training, one can use the summaries obtained from the encoder network directly for downstream tasks, such image classification, object detection, and, as we will show in the following sections, parameter inference. 
The network used for the downstream task tends to have a simpler architecture than the encoder network, such as a multi-layer perceptron with a few dense layers. 
This offers another potential advantage of self-supervised methods: once the encoder model has been trained, training the network on the summaries for downstream tasks is usually faster and more efficient than training a supervised model directly on the input data. Furthermore, self-supervised learning methods have been empirically and theoretically shown to generalize better to out-of-distribution data, which could be partially attributed to the simpler structure of the neural network specialized for the downstream task \citep{SSL_generalization_gap}.

A key difficulty of implementing self-supervised methods is a phenomenon called \textit{collapse}, in which the encoder network learns a trivial solution and produces the same summaries for different input vectors \citep{SSL_review,SSL_cookbook}. A variety of approaches have been introduced in order to deal with this problem and enable learning meaningful representations.
In this work, we focus on a particular approach called VICReg (Variance-Invariance-Covariance Regularization) \citep{bardes2021vicreg}, described in the next subsection. 
VICReg is designed to explicitly avoid the collapse problem through an easily interpretable \textit{triple objective function}, which maximizes the similarity of the summaries corresponding to the same image, while minimizing the redundancy between different features of the summary vectors and maintaining variance between summaries within a training batch. 

A complementary approach to the collapse problem involves \emph{contrastive learning} methods, which separate the training samples into `positive' and `negative' pairs (e.g. \citet{SSL_contrastive,SSL_contrastive_triplet,SSL_contrastive_n_pair,SSL_MoCo,chen2020simple}). These pairs then contribute adversarially to the overall objective function in the pre-training step: the loss encourages the encoder to create similar summaries for the `positive' pairs, while pushing the summaries for the `negative' pairs apart in representation space. For completeness, we also test the self-supervised learning approach with a canonical contrastive learning method called SimCLR \citep{chen2020simple} and find comparable performance to the VICReg baseline. In Appendix \ref{appendix:SimCLR_vs_VICReg}, we provide a brief summary of SimCLR and a more detailed overview of our implementation and results. We emphasize that our method is not dependent on a specific choice of self-supervision method, since it relies more generally on the paradigm of self-supervision.
\subsection{
Variance-Invariance-Covariance Regularization (VICReg)}\label{sec:SSL_VICReg}

The VICReg framework was introduced and described in \cite{bardes2021vicreg}. In this section, we briefly review its key aspects as well as extensions introduced to make it applicable to specific use cases in cosmology. 

Similarly to other self-supervised methods, VICReg can be divided into a pre-training step and a downstream task. During the pre-training step, the encoder network is first provided with two different \textit{views} $X$ and $X'$ of an input $I$. In the image domain, so-called \textit{views} are random transformations of the image $I$ obtained by, for instance, cropping it at different locations, applying color jitters or blurring the image. In the context of cosmological data, different views might represent different realizations of an observable that corresponds to the same fundamental cosmological parameters, but, for instance, different initial conditions or evolution histories.

The encoder uses views $X$ and $X'$ to produce corresponding low-dimensional summaries $S$ and $S'$. The summaries are then used as an input to a small expander network that maps them onto vectors $Z$ and $Z'$, called \textit{embeddings}. 

\begin{figure*}
\centering
 \includegraphics[width=\textwidth]{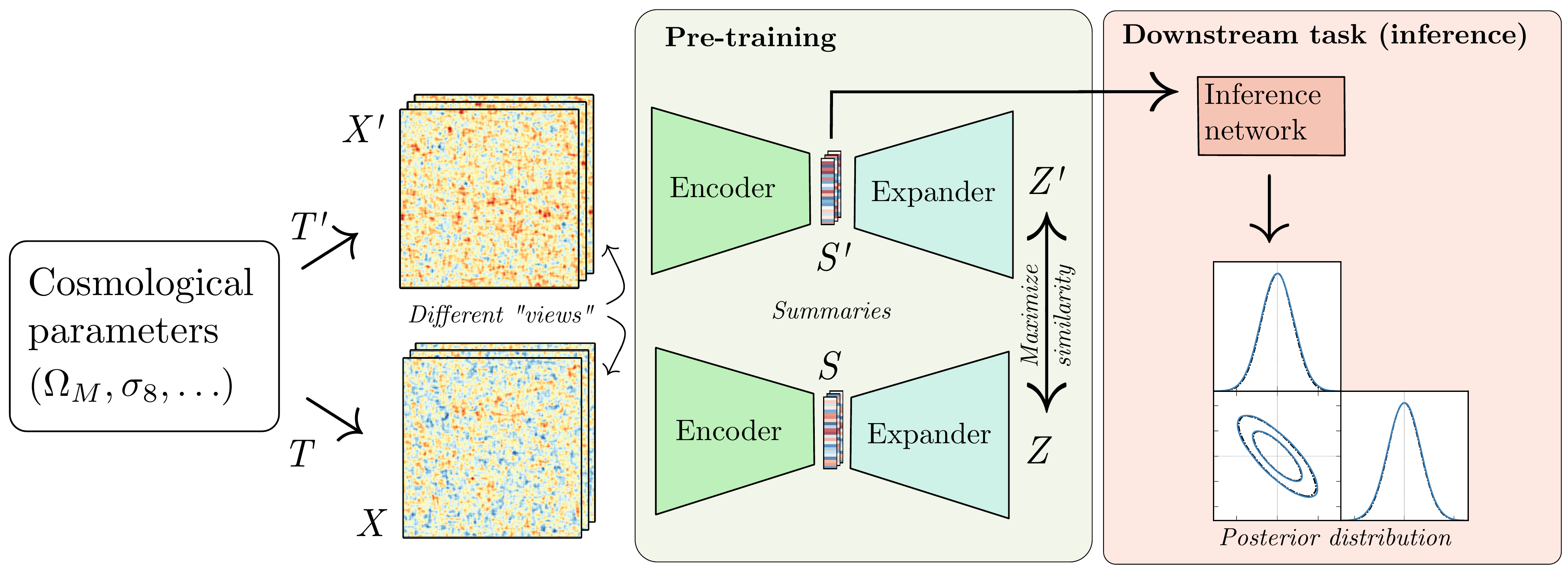}
 \caption{A schematic overview of the self-supervised learning pipeline implemented in this work. 
The $T, T'$ are different transformations used to produce two views (e.g., lognormal density maps) $X, X'$ of the same underlying cosmological parameters of interest (e.g., $\Omega_M$ and $\sigma_8$). 
The inference network is trained on the summaries $S, S'$ obtained from the pre-training step.}
 \label{fig:pipeline}
\end{figure*}

Empirically, it has been found that computing self-supervised losses on embeddings $Z, Z'$ results in more informative summaries than computing the loss directly on the summaries $S, S'$ (e.g. \citet{chen2020improvedMoCo,chen2020simple,zbontar2021barlow,bardes2021vicreg}). Although the expander network is discarded after the pre-training step, using the expander network generally results in substantial improvement of the performance of the summaries $S, S'$ on the downstream tasks. This behaviour is most likely due to the fact that, by applying a non-linear transformation to the summaries $S, S'$, the encoder network can act as a `filter', which removes features from the representations $Z, Z'$ \citep{chen2020simple} that could be useful later on in the downstream task \citep{SSL_ProjectionHead_goodPractices_appalaraju2020,SSL_usefulnessProjectionHead_gupta2022}. These features are not particularly useful or important for the self-supervised loss functions, but they could be leveraged later on, when using the summaries for the downstream tasks. In this manner, the expander network allows for more flexible and informative summaries of the input images. Therefore the VICReg loss is computed on the level of embeddings $Z, Z'$ , but the summaries $S, S'$ are used for the downstream tasks in the subsequent steps of the method. We show a schematic overview of the method in Fig. \ref{fig:pipeline}.

Let $Z = [Z_1, ..., Z_n], \, \mathrm{and} \, Z^{\prime} = [Z_1^{\prime}, ..., Z_n^{\prime}]$ be two batches of $n$ embeddings, where each embedding $Z_i$ is a $d$-dimensional vector. The three terms of the VICReg objective function are then defined as follows.

The invariance component $s$ measures the similarity between the outputs of the encoder and is computed as the mean-squared Euclidean distance between all pairs of embeddings in $Z, \, Z^{\prime}$:
\begin{equation}\label{eq:invariance_loss}
s(Z, Z^{\prime})=\frac{1}{n} \sum_{i=1}^{n}\left\|Z_i-Z_i^{\prime}\right\|_2^2
\end{equation}

The variance loss $v$ is intended to prevent \textit{norm collapse} which occurs when the encoder maps every input to the same (trivial) output. It measures the overall variance in a given batch across different dimensions in the embedding space and encourages the variance along each dimension $j$ to be close to some constant $\gamma$. Let $Z^j$ be a vector that consists of the values of the embeddings $Z_i$ at $j$-th dimension. Then the variance loss can be computed as:
\begin{equation}\label{eq:variance_loss}
v(Z)=\frac{1}{d} \sum_{j=1}^d \max \left(0, \gamma-S\left(Z^j, \epsilon\right)\right),
\end{equation}
where $S(x, \epsilon)=\sqrt{\operatorname{Var}(x)+\epsilon}$ is defined as the standard deviation which is regularized by a small scalar $\epsilon$. Following \cite{bardes2021vicreg}, in our implementation $\gamma$ and $\epsilon$ are fixed to 1 and $10^{-4}$ respectively.

The covariance loss $c(Z)$ is used to address the \textit{informational collapse} whereby different dimensions of the summaries encode the same information and are therefore redundant. It drives the covariance matrix $\mathbb{C}(Z)$ to be close to a diagonal matrix by minimizing the sum of the squares of the off-diagonal entries of the covariance matrix: \begin{equation}\label{eq:covariance_loss}
c(Z)=\frac{1}{d} \sum_{k \neq l}[\mathbb{C}(Z)]_{k, l}^2.
\end{equation}
The covariance matrix $\mathbb{C}(Z)$ is defined as:
\begin{equation}\label{eq:covariance_loss_C}
\mathbb{C}(Z)=\frac{1}{n-1} \sum_{i=1}^n\left(Z_i-\bar{Z}\right)\left(Z_i-\bar{Z}\right)^T, \, \, \mathrm{where} \, \bar{Z} = \frac{1}{n}\sum_{i=1}^{n}Z_i.
\end{equation}

The final loss function is a weighted sum of invariance $s$, variance $v$ and covariance $c$ terms: 
\begin{equation}\label{eq:VICReg_loss_function}
\ell(Z, Z^{\prime}) = 
\lambda s(Z, Z^{\prime})+\mu[v(Z)+v(Z^{\prime})] +\nu[c(Z)+c(Z^{\prime})],
\end{equation}
where $\lambda, \, \mu, \, \nu$ are hyperparameters controlling the weights assigned to each term in the loss function.
\subsection{Downstream Task: Parameter Inference}

After pre-training, the summaries can be used for downstream tasks by training a simple neural network, such as a multi-layer perceptron with a few layers, on the task. We use the summaries to infer cosmological parameters of interest and refer to the neural network used in this step as the \textit{inference network}. Assuming a Gaussian likelihood, we use the inference network to predict the parameters' means $\theta_n$ and covariances $\Sigma_n$  by minimizing the negative log-likelihood function:
\begin{equation}\label{eq:Regression_Loss}
\mathcal{L} = \frac{1}{N} \sum_{n=1}^{N} \left[\frac{1}{2} \ln |\Sigma_n| + \frac{1}{2} \left(\theta_n-\mu_n\right)^T \Sigma^{-1}_n\left(\theta_n -\mu_n\right)\right],
\end{equation}
where $\mu_n$ are the true values of the parameters, $\theta_n$ and $\Sigma_n$ are the predicted means and covariance matrix of the parameters, and the sum runs over all input images in the training set. 

We emphasize that, even though we showcase the specific downstream task of parameter inference, representative summaries can be constructed for a wide variety of downstream tasks common in cosmology and astrophysics. 
For instance, sensitivity analyses of simulation-based inference typically focus on a particular summary statistics in order to examine the robustness of the inference to different components of cosmological forward models. Summary statistics constructed in this study could be considered as an alternative to the traditional summary statistics (such as power spectrum or bispectrum) used for such sensitivity analyses \citep{modi2023sensitivity}. 
Beyond parameter inference, summaries can be constructed for e.g. source identification and deblending \citep{2021arXiv210202409L,2022mla..confE..27H}, allowing for massive compression of survey data while retaining the desired information.

\section{Self-Supervision for Data Compression}\label{sec:VICreg_data_compression}

\subsection{Lognormal Fields}\label{sec:LognormalFields}

We first test our methodology on mock cosmological data: lognormal random fields generated from linear matter power spectra. Lognormal fields are commonly used as a rough approximate description of matter density fields \citep{Percival_2004_LN_fields_galaxy_clustering,Beutler_2011_LN_fields_mock_galaxy_catalogs,Cole_2005_LN_fields_mock_galaxy_catalogs}. 
While they cannot accurately capture small-scale features of the cosmological density fields (e.g. \citet{Kitaura_2010_LN_fields_density_reconstruction}), they nevertheless serve as a useful model for large-scale structure due to a number of properties, including the ability to compute summaries and information content analytically \citep{Coles_Jones_1991_LN_fields_theory}. Unlike Gaussian density fields,  lognormal density fields take positive values by construction, and they have been shown to agree with the results of $N$-body simulations even in mildly non-linear regimes \citep{Kayo_LN_fields_Nbody_sims}. 

\subsubsection{Data}\label{sec:LognormalFields_Data} 

We generate lognormal fields $\delta_{LN}(x)$ from 2D Gaussian overdensity fields $\delta_{G}(x)$ with a specified power spectrum $P_{G}(k)$. We convert the Gaussian fields to obtain corresponding lognormal overdensity fields: 

\begin{equation}\label{eq:LN_fields_from_Gaussian_fields}
    \delta_{LN}(x) = \exp{\left(\delta_{G}(x) - \frac{1}{2}\sigma_{G}^{2}\right)} - 1,
\end{equation}
where $\sigma_{G}^{2}$ is the variance of the field $\delta_{G}(x)$. The Gaussian fields are produced with the \texttt{powerbox} package \citep{Murray2018_software_powerbox}.

\begin{figure*}
\begin{subfigure}{0.33\textwidth}
    \centering
    \includegraphics[width = \textwidth,]{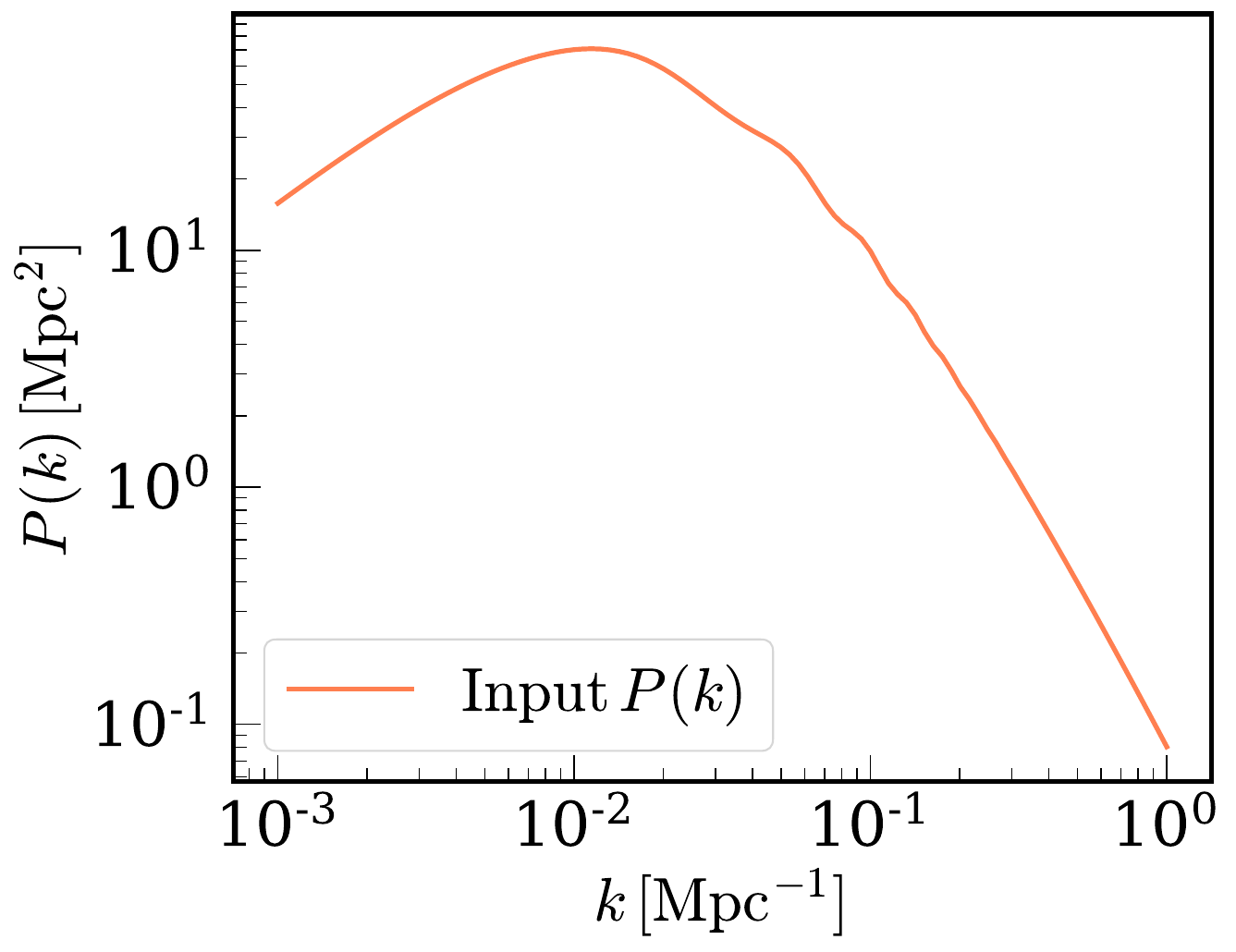}
\end{subfigure} 
\begin{subfigure}{0.61\textwidth}
    \centering
    \includegraphics[width=\textwidth,]{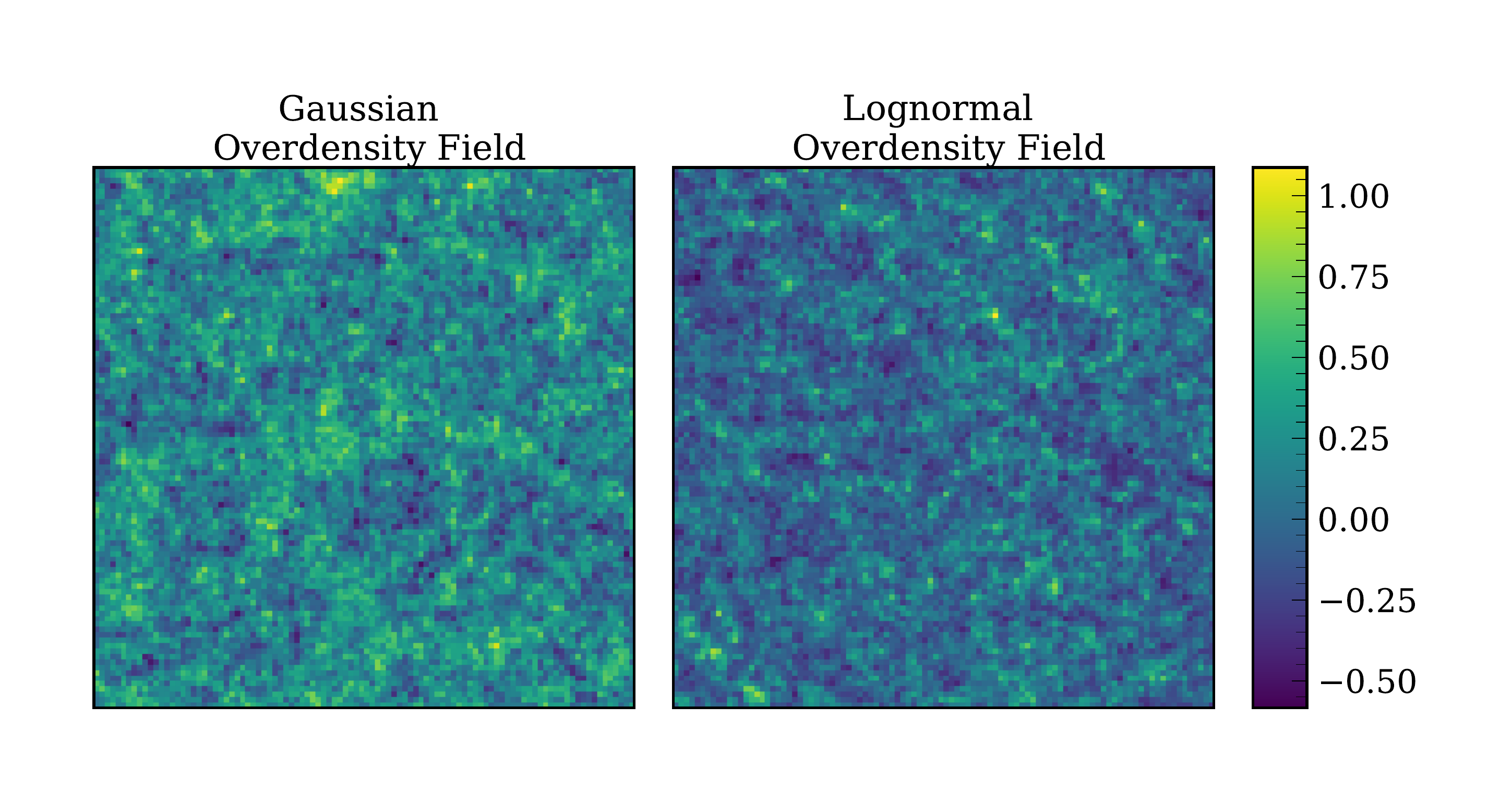}
\end{subfigure} 
\caption{An example of a Gaussian overdensity field $\delta_G$ and a corresponding lognormal overdensity field $\delta_{LN}$ generated from a linear matter power spectrum $P_G$ with Eisenstein-Hu transfer function and $\Omega_M = 0.3, \sigma_8 = 0.8$.} 
\label{fig:LN_fields_example}
\end{figure*}

We take $P_{G}(k)$ to be the linear matter power spectrum computed with the Eisenstein-Hu transfer function \citep{Eisenstein_Hu_LN_fields_Pk} and generate the power spectra using the \texttt{pyccl} package \citep{Chisari_software_ccl}. 
For each $P_{G}(k)$, we vary two cosmological parameters: total matter density, $\Omega_M$, and the r.m.s. of the present day ($z$ = 0) density perturbations at scales of 8 $h^{-1}$ Mpc, $\sigma_8$.  
We fix the remaining cosmological parameters to the following values: $\Omega_{b} = 0.05, \, h = 0.7, \, n_s = 0.96, N_{\mathrm{eff}} = 3.046, \sum m_{\nu} = 0$ eV. 
We use a grid of $N^{2} = 100\times100$ points and set the area of the slice to be $A = L^{2} = (1000 \, \mathrm{Mpc})^{2}$. Figure \ref{fig:LN_fields_example} shows an example of a power spectrum $P_{G}$ with $\Omega_M = 0.3$ and $\sigma_8 = 0.8$ as well as the corresponding realizations of Gaussian and lognormal overdensity fields. 

We generate a set of 10,000 different combinations of cosmological parameters uniformly distributed in the range $\Omega_M \in [0.15, 0.45]$ and $\sigma_8 \in [0.65, 0.95]$. 
For each combination of $\Omega_M$ and $\sigma_8$, we simulate 10 different realizations of lognormal overdensity fields, constructed from different initial random seeds. 
These realizations, rotated and flipped at random, serve as augmentations (`different views') to train the VICReg encoder network. 

\subsubsection{VICReg Setup}\label{sec:LognormalFields_VICReg}

We compress the $100\times100$ dimensional maps down to 16-dimensional summaries using an encoder network with 9 convolutional layers and 2 fully-connected layers. The inference network used to infer parameters from the compressed summaries is a simple fully-connected neural network with 2 hidden layers. We provide the details about the architectures of the two networks in Appendix \ref{appendix:LN_fields_NN_Info}.

We use 80\% of the data for training, 10\% for validation, and the remaining 10\% for testing. When splitting the dataset into the training, validation, and testing sets, augmentations corresponding to the same set of parameters are not shared across the different data splits. 

We train the encoder network on the training set for 200 epochs in the \texttt{PyTorch} \citep{PytorchDLL} framework using \texttt{AdamW} \citep{kingma2014adam,hutter2019AdamW} optimizer, which is used as a default optimizer in this work, with initial learning rate of $2\times10^{-4}$ and cosine annealing schedule. Throughout the work we also perform a manual hyperparameter search to find the optimal invariance $\lambda$, variance $\mu$, and covariance $\nu$ weights in the loss function. We work with $\lambda=5$, $\mu=5$, and $\nu=1$. 

The downstream network is trained for 200 epochs with initial learning rate $10^{-3}$, reduced by a factor of 5 when the validation loss plateaus for 10 epochs.

In this work, we use the same training, validation, and test datasets when training the encoder network and the downstream inference network. We evaluate the performance of neural networks on the validation set at the end of each epoch and save the models with best validation loss. Once both the networks are trained, the overall performance of the algorithm is evaluated on the test dataset. 

As a baseline, we also construct a convolutional neural network with an architecture that is equivalent to a combination of the encoder and inference networks, corresponding to the fully-supervised case. 
We train this network to perform field-level inference without intermediate steps: given an overdensity map, it learns to infer the means and covariance matrix of the cosmological parameters $\Omega_M$ and $\sigma_8$. The network is trained for 200 epochs, with initial learning rate $2\times10^{-3}$ and cosine annealing schedule. 
This model is used to evaluate the performance of the self-supervised method compared to a fully-supervised benchmark. 

\subsubsection{Results}\label{sec:LognormalFields_Results}
\begin{figure*}
\centering
 \includegraphics[width=\textwidth]{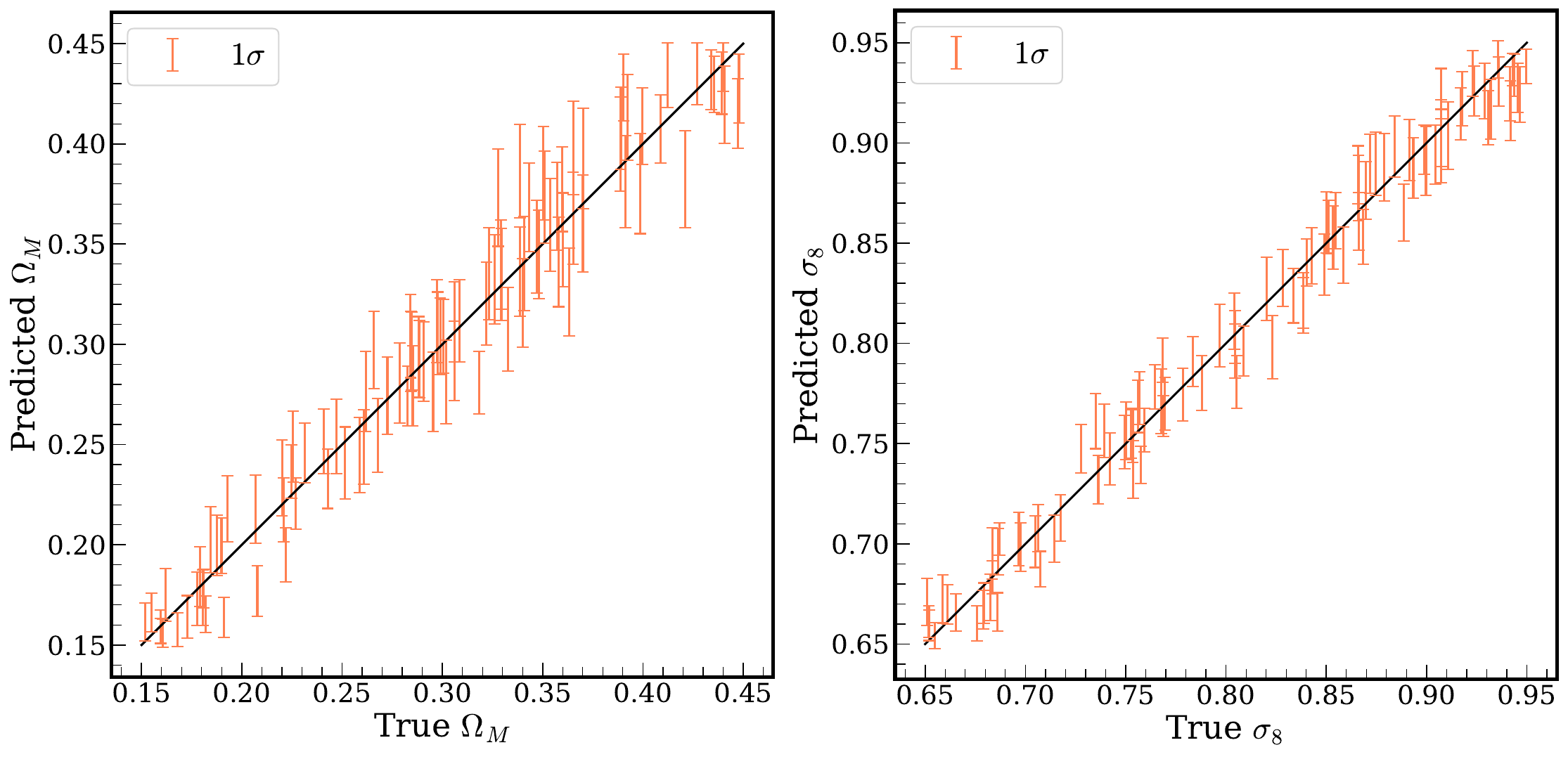}
 \caption{Scatter plot of predicted means and 1-$\sigma$ uncertainties of cosmological parameters $\Omega_M$ (left) and $\sigma_8$ (right) plotted against the true values of the parameters for a subset of a 100 maps from the test set. Predictions for the means and variances of the parameters were obtained by training a simple inference neural network on VICReg summaries.}
 \label{fig:LN_fields_truth_vs_prediction}
\end{figure*}

We now present the results of the parameter inference with VICReg method on a test dataset of 10,000 mock overdensity maps. In Fig. \ref{fig:LN_fields_truth_vs_prediction}, we plot the predicted values of $\Omega_M$ (left panel) and $\sigma_8$ (right panel) compared to the true values of these parameters for 100 test maps, with the error bars showing the $1\sigma$ uncertainties predicted by the model. 

We compare the performance of the VICReg method to the performance of the baseline supervised learning method on the test dataset in Table \ref{tab:compare_VICReg_CNN}. 
We find that the inference network trained on VICReg summaries is able to recover the true values of cosmological parameters with both accuracy and precision, with relative errors on $\Omega_M$ and $\sigma_8$ equal to $5.2\%$ and $1.3\%$, respectively. 
For comparison, a neural network with an equivalent architecture, trained on the maps directly in a fully-supervised manner, predicts the cosmological parameters with similar accuracy (relative errors on $\Omega_M$ and $\sigma_8$ are equal to $5.1\%$ and $1.3\%$), which suggests that the encoder network has learned an effective compression scheme that reduces the maps to summaries without substantial loss of information.

\begin{table}
 \centering
 \begin{tabular}{c c c c c}
  \hline
  \bf{Method} & \bf{Loss} & \bf{MSE} & \bf{MSE} on $\Omega_M$ & \bf{MSE} on $\sigma_8$ \\
   & & & (Relative error) & (Relative error) \\ \hline
  VICReg & -5.98 & 2.7$\times 10^{-4}$ & 3.6$\times 10^{-4}$  & 1.8$\times 10^{-4}$ \\
    & & & (5.2\%) & (1.3\%) \\ \hline
  Supervised & -6.01  & 2.6$\times 10^{-4}$ & 3.4$\times 10^{-4}$ & 1.7$\times 10^{-4}$ \\
  & & & (5.1\%) & (1.3\%) \\
  \hline
 \end{tabular}
 \caption{Summary of the performance of the VICReg method and of an equivalent supervised baseline model for inferring cosmological parameters $\Omega_M$ and $\sigma_8$ from lognormal overdensity maps. The performance of the two methods is evaluated on the test dataset described in Sec. \ref{sec:LognormalFields_Data}.}
 \label{tab:compare_VICReg_CNN}
\end{table}

\subsubsection{Comparison to Theoretical Expectation}\label{sec:LognormalFields_Fisher}

As a bijective transformation from a Gaussian random field, the lognormal field preserves the information content of an equivalent Gaussian one and is therefore conducive to an analytic treatment of its information content. We use this fact to compare expected constraints given the Fisher information content of the underlying lognormal field and VICReg-extracted summaries. 

The Fisher information matrix of the Gaussian fields can be conveniently computed from their power spectra $P_{G}$. The elements of the Fisher matrix for parameters $\theta_{\alpha}, \theta_{\beta}$ are given by: 
\begin{equation}
    F_{\alpha \beta}= \frac{1}{2}\sum_{k} \frac{\partial P_{G}(k)}{\partial \theta_{\alpha}} \frac{\partial P_{G}(k)}{\partial \theta_{\beta}} \frac{1}{P_{G}(k)^2},
\end{equation}
where the sum is over all independent $k$-modes. The Fisher matrix elements for the overdensity maps are computed by evaluating the linear matter power spectrum $P_G$ at the relevant $k$-values and numerically computing the derivatives of the power spectrum with four-point finite differences formula.

Assuming the summaries $S$ we obtain with VICReg can be described by a Gaussian likelihood, we can compute the Fisher information matrix for $S$ as follows:
\begin{equation}\label{eq:Fisher_matrix_summaries}
F_{\alpha \beta}=\frac{\partial S}{\partial \theta_\alpha} C^{-1} \frac{\partial S}{\partial \theta_\beta}.
\end{equation}
We evaluate the Fisher matrix for the summaries numerically: the derivative of the summaries are computed using four-point finite differences formula, and the covariance matrix $C$ is estimated with Ledoit-Wolf method implemented in the \texttt{sklearn} package. 
We use 10,000 realizations of lognormal maps at the fiducial cosmology to evaluate the covariance matrix $C$ and 1,000 realizations of lognormal maps around the fiducial parameters to compute the derivatives $\partial S/ \partial \theta_\alpha$. 

One can then use the Cramer-Rao bound to estimate the minimum variance of the parameter of interest $\theta_{\alpha}$ as the inverse of the Fisher matrix:
\begin{equation}\label{eq:CramerRaoBound}
    \sigma_{\alpha} \geq [F^{-1/2}]_{\alpha \alpha}.
\end{equation}

\begin{figure}
 \includegraphics[width=\columnwidth]{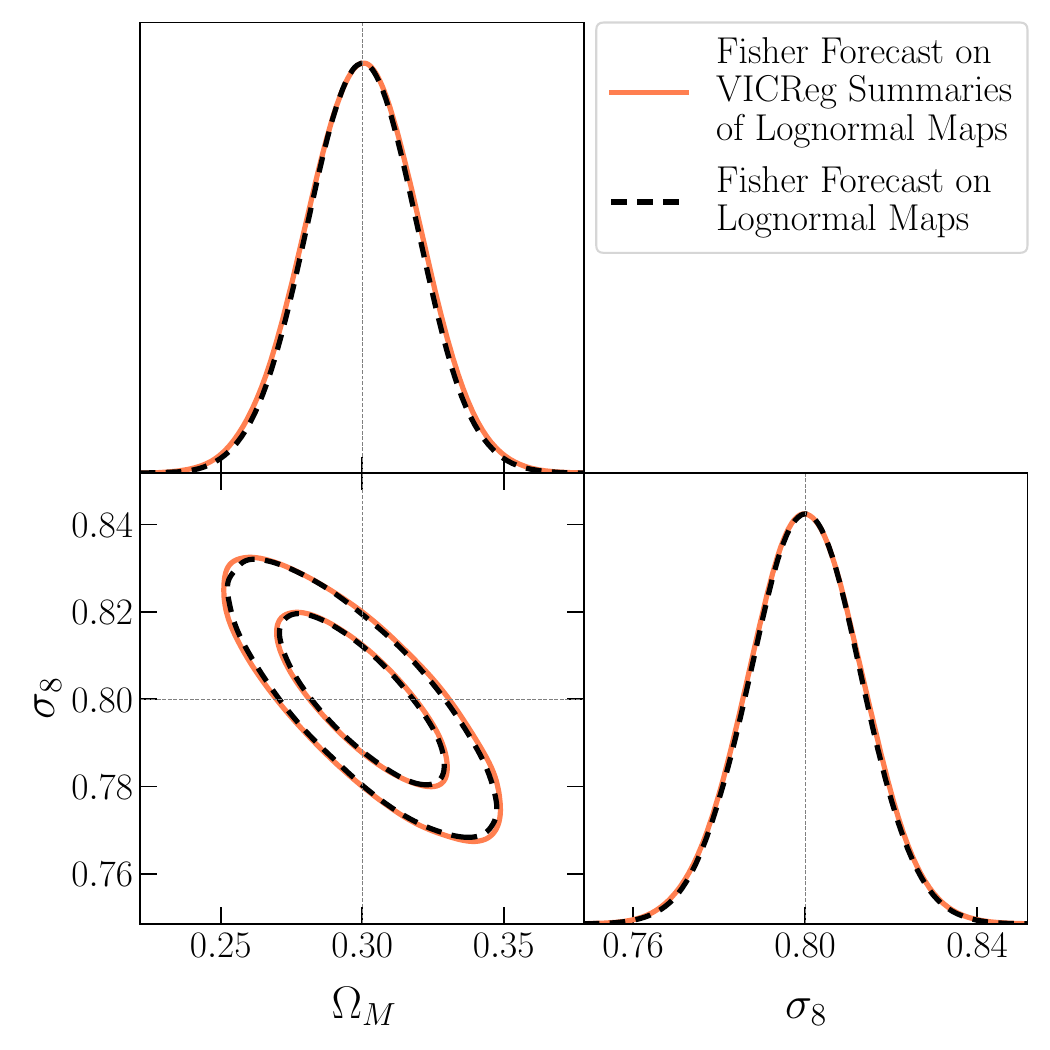}
 \caption{Constraints from Fisher forecast on the cosmological parameters $\Omega_M$ and $\sigma_8$ obtained from lognormal overdensity maps (black dash-dotted line) and from summaries constructed with VICReg (orange solid line). The results shown on the plot were obtained for a fiducial cosmology with $\Omega_M = 0.3$ and $\sigma_8 = 0.8$.}
 \label{fig:LN_fields_fisher_forecast}
\end{figure}

In Fig. \ref{fig:LN_fields_fisher_forecast}, we show the Fisher-informed constraints on $\Omega_M$ and $\sigma_8$. The fiducial values of $\Omega_M$ and $\sigma_8$ in this case are set to 0.3 and 0.8 respectively. The Fisher-informed contours from the lognormal fields and the VICReg summaries are in excellent agreement, demonstrating that the summaries preserve the Fisher information content of the maps almost entirely. We show the lower bounds on the errors on the two parameters in Table \ref{tab:compare_summaries_Fisher}. 
Since the Fisher-informed constraints from the VICReg summaries of the lognormal overdensity maps were computed under a set of assumptions about Gaussianity, we further examine and validate our conclusions in Appendix \ref{appendix:Fisher_forecast_examination}. We compare the Fisher-informed constraints to posterior distributions inferred by a normalizing flow trained on the VICReg summaries and find that similar conclusions hold.

\begin{table}
 \centering
 \begin{tabular}{c c c}
  \hline
  \bf{Data} & \bf{Fisher Error on $\Omega_M$} & \bf{Fisher Error on $\sigma_8$}\\
   & (Relative error) & (Relative error) \\ 
  \hline
  Maps & 0.019 & 0.013 \\
  & (6.4\%) & (1.6\%) \\ \hline
  VICReg Summaries & 0.020 & 0.013 \\
  & (6.7\%) & (1.7\%) \\
  \hline
 \end{tabular}
 \caption{Summary of the Fisher-informed constraints on $\Omega_M$ and $\sigma_8$ obtained from lognormal overdensity maps and from VICReg summaries for the fiducial cosmology with $\Omega_M = 0.3$ and $\sigma_8 = 0.8$.}
 \label{tab:compare_summaries_Fisher}
\end{table}

\subsection{CAMELS Total Matter Density Maps}\label{sec:CAMELS}

Having demonstrated the potential of self-supervised learning for data compression and parameter inference on a simple lognormal model of overdensity fields, we now consider an application of our method to more realistic data: total matter density maps from the CAMELS project \citep{CMD,CAMELS}. CAMELS is a collection of hydrodynamic and $N$-body simulations, which span a wide range of cosmological ($\Omega_M, \, \sigma_8$) and astrophysical parameters (stellar feedback parameters $A_{\mathrm{SN} 1}, A_{\mathrm{SN} 2}$ and AGN feedback parameters $A_{\mathrm{AGN} 1}, A_{\mathrm{AGN} 2}$). Stellar feedback parameters $A_{\mathrm{SN} 1}, A_{\mathrm{SN} 2}$ parametrize the galactic stellar-driven winds or outflows which eject the gas from the interstellar medium to large distances away from the star-forming galaxy. AGN parameters $A_{\mathrm{AGN} 1}, A_{\mathrm{AGN} 2}$ describe the feedback from the massive black holes, which affects the large-scale matter distribution by heating up and expelling the gas from the galaxy \citep{baryons_feedback_review_somerville_dave_2015}. We refer the readers to \citet{CAMELS} for further details on the CAMELS dataset.

In this work, we use two publicly available suites of hydrodynamic CAMELS simulations, which implement two distinct galaxy formation models: IllustrisTNG \citep{IllustisTNG_sims_1,IllustisTNG_sims_2} and SIMBA \citep{SIMBA_simulations}. We use the latin hypercube (LH) sets of the two suites, which contain realizations from uniformly-drawn cosmological parameters ($\Omega_M \in [0.1,0.5]$, $\sigma_8 \in[0.6, 1.0]$) and astrophysical parameters. Each simulation in the LH sets has a different value of the random seed which defines the initial conditions for the simulation. For our study, we use the total matter density maps from the CAMELS multifield dataset, which represent spatial distribution of baryonic as well as dark matter at $z=0$ within a $25 \times 25 \times 5 \, (h^{-1}\mathrm{Mpc})^3$ slice of a given simulation \citep{CMD}. IllustrisTNG and SIMBA datasets contains 15,000 different maps each (1000 hydrodynamic simulations with 15 maps per simulation). We construct self-supervised summaries using these maps and demonstrate their efficacy for downstream parameter inference. 

\subsubsection{VICReg Setup}\label{sec:CAMELS_data_vicreg}

We modify the notion of two different views/augmentation to represent total mass density maps from two different slices of the same simulation, rotated or flipped at random during training. This should enable the encoder network to learn relevant cosmological information from the maps and become insensitive to spatial variations in the slices.

We also find it helpful to modify the VICReg loss such that each batch includes 5 pairs of different augmentations from each simulation, as opposed to a single pair per simulation (or per set of cosmological parameters). Since the CAMELS maps have more complexity than the lognormal maps, this allows the encoder network to learn from more variations. 

Due to the high computational cost of running hydrodynamic simulations, IllustrisTNG and SIMBA have fewer data samples than the lognormal maps dataset used in Sec. \ref{sec:LognormalFields}, so we reserve more data for validation and testing purposes: 70\% of the simulations for training, 20\% for validation, and the remaining 10\% for testing. 

We use ResNet-18 \citep{ResNet_model, PytorchDLL} as the encoder, which compresses the $256 \times 256$ maps to summaries of length 128. The inference network used for parameter inference is a simple fully-connected 2-layer neural network, with 512 units in each layer. 

We train the encoder for 150 epochs with initial learning rate  $10^{-3}$, which is multiplied by a factor of 0.3 when the validation loss plateaus for 10 epochs. The weights $\lambda, \, \mu, \, \nu$ in the loss function are set to 25, 25, and 1, respectively. The inference network is trained with initial learning rate $7 \times 10^{-4}$. 

As a baseline model to compare against, we train a ResNet-18 model in a fully supervised manner to infer $\Omega_M, \, \sigma_8$ directly from the total mass density maps. We train the network for 200 epochs with initial learning rate of $2\times 10^{-4}$ and cosine annealing schedule. 

\subsubsection{Results}\label{sec:CAMELS_results}

\begin{figure*}
\begin{subfigure}{\textwidth}
    \centering
    \includegraphics[width=\textwidth]{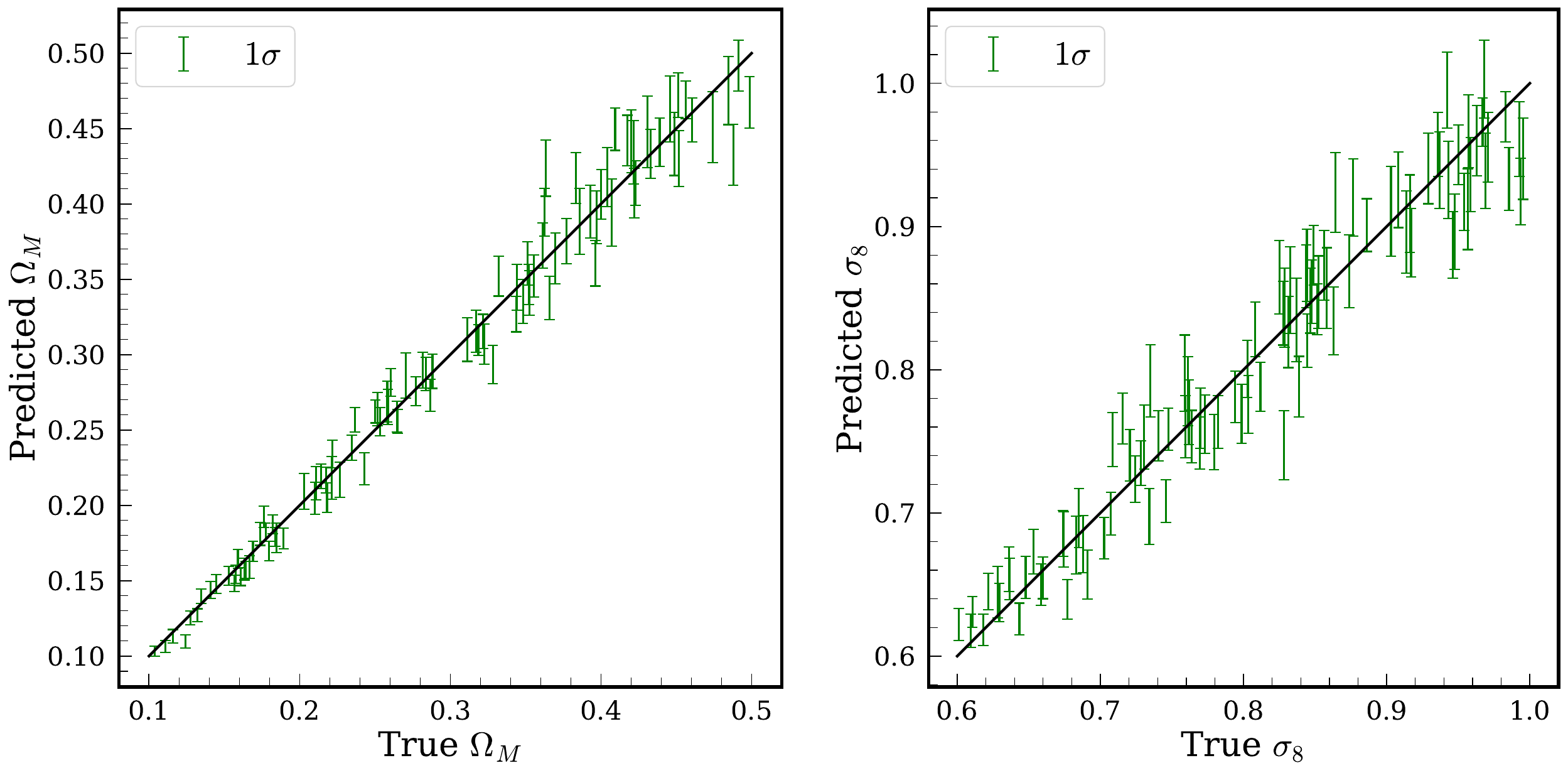}
    \caption{VICReg: SIMBA.}
    \label{fig:SIMBA_fields_truth_vs_prediction}
\end{subfigure}
\begin{subfigure}{\textwidth}
    \centering
    \includegraphics[width=\textwidth]{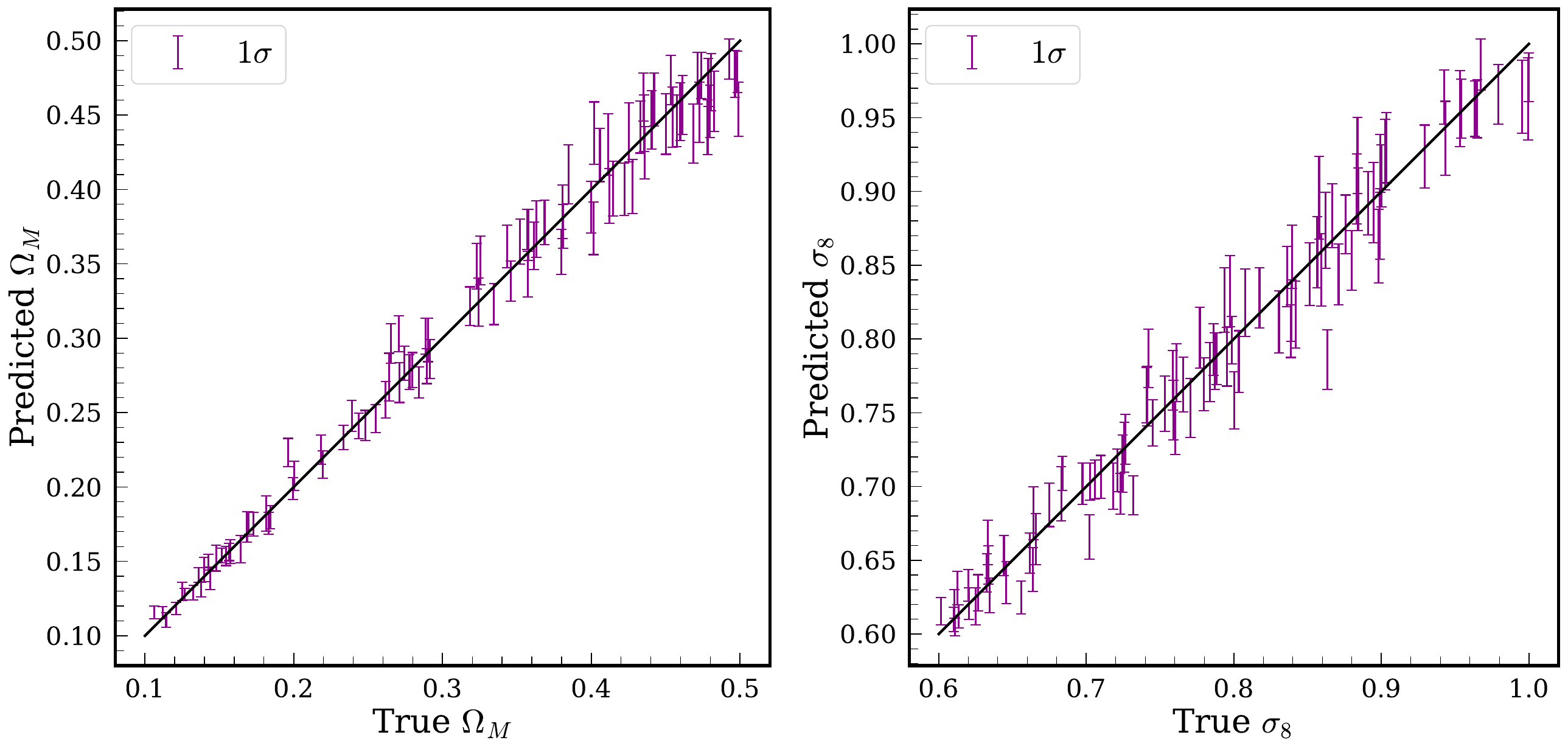}
    \caption{VICReg: IllustrisTNG.}
    \label{fig:IllustrisTNG_fields_truth_vs_prediction}
\end{subfigure}       
\caption{Predicted means and 1-$\sigma$ uncertainties of cosmological parameters $\Omega_M$ and $\sigma_8$ compared to the true values of the parameters for total matter density maps from SIMBA and IllustrisTNG simulations. Predictions for the means and variances of the parameters were obtained by training simple inference neural network on VICReg summaries.}
\label{fig:SIMBA_IllustrisTNG_fields_truth_vs_prediction}
\end{figure*}

We next present our results for the two simulations suites. We plot the predicted values of $\Omega_M$ (left panel) and $\sigma_8$ (right panel) against the true values for a subset of maps from the test set for the SIMBA (Fig. \ref{fig:SIMBA_fields_truth_vs_prediction}) and IllustrisTNG (Fig. \ref{fig:IllustrisTNG_fields_truth_vs_prediction}) simulation suites. The error bars on the plots correspond to the predicted $1\sigma$ uncertainties. 

We summarize the errors on the predicted parameters and compare the performance of the VICReg algorithm on the two simulation suites in Tables \ref{tab:compare_VICReg_CNN_SIMBA} and \ref{tab:compare_VICReg_CNN_IllustrisTNG}. It can be seen that the inferred parameters provide a fairly accurate and unbiased estimate of the true parameters. Trained directly on the VICReg summaries, the inference model is able to infer the cosmological parameters with percent-level accuracy: the relative errors on $\Omega_M$ and $\sigma_8$ are $3.8\%$ and $2.5\%$ respectively for the SIMBA suite, and $3.7\%$ and $1.9\%$ for the IllustrisTNG suite. 

We find that performing field-level inference on the matter density maps with an equivalent (ResNet-18) supervised model results in similar constraints on the cosmological parameters: the relative errors on $\Omega_M$ and $\sigma_8$ are $3.3\%$ and $2.3\%$ respectively for the SIMBA suite and $3.3\%$ and $1.8\%$ for the IllustrisTNG suite. These results suggest that, despite massive reduction in the size and dimensionality of the data, the VICReg encoder network learns a near-optimal compression scheme with only a slight reduction in sensitivity of downstream inference.

\begin{table}
\begin{subtable}[h]{\columnwidth}
 \centering
 \begin{tabular}{c c c c c}
  \hline
  \bf{Method} & \bf{Loss} & \bf{MSE} & \bf{MSE on $\Omega_M$} & \bf{MSE on $\sigma_8$} \\
   & & & (Relative error) & (Relative error) \\ \hline
  VICReg & -3.27 & 4.24$\times 10^{-4}$ & 2.09$\times 10^{-4}$  & 6.37$\times 10^{-4}$ \\
    & & & (3.79\%) & (2.46\%) \\ \hline
  Supervised & -3.61  & 3.78$\times 10^{-4}$ & 1.80$\times 10^{-4}$ & 5.75$\times 10^{-4}$ \\
  & & & (3.28\%) & (2.29\%) \\
  \hline
 \end{tabular}
 \caption{SIMBA.}
 \label{tab:compare_VICReg_CNN_SIMBA} 
\end{subtable}

\begin{subtable}[h]{\columnwidth}
\begin{tabular}{c c c c c}
  \hline
  \bf{Method} & \bf{Loss} & \bf{MSE} & \bf{MSE on $\Omega_M$} & \bf{MSE on $\sigma_8$} \\
   & & & (Relative error) & (Relative error) \\ \hline
  VICReg & -3.60 & 3.13$\times 10^{-4}$ & 2.48$\times 10^{-4}$  & 3.79$\times 10^{-4}$ \\
    & & & (3.70\%) & (1.89\%) \\ \hline
  Supervised & -3.84  & 2.71$\times 10^{-4}$ & 1.97$\times 10^{-4}$ & 3.46$\times 10^{-4}$ \\
  & & & (3.32\%) & (1.83\%) \\
  \hline
 \end{tabular}
 \caption{IllustrisTNG.}
 \label{tab:compare_VICReg_CNN_IllustrisTNG} 
 \end{subtable}
 \caption{Summary of the performance of the VICReg method and of an equivalent supervised model for inferring cosmological parameters $\Omega_M$ and $\sigma_8$ from SIMBA  and IllustrisTNG total matter density maps, evaluated on the respective test datasets.}
\end{table}

\subsubsection{Performance on an Out-of-Distribution Dataset} \label{sec:CAMELS_VICReg_out_of_distribution}
When testing the models on the out-of-distribution data, we find that similar results hold. 
We summarize the performance of the VICReg method and compare it to the supervised baseline model in Tables \ref{tab:app_compare_VICReg_CNN_SIMBA_on_I} and \ref{tab:app_compare_VICReg_CNN_IllustrisTNG_on_S} for two scenarios of out-of-distributions datasets: applying models trained on data from IllustrisTNG simulations to total matter density maps from the SIMBA suite, and applying applying models trained on data from SIMBA simulations to total matter density maps from the IllustrisTNG suite. We find that, similarly to the case with both training and test data coming from the same distribution, the supervised baseline model shows slightly better performance than the inference network trained on the summaries of the maps.

\begin{figure*}
\begin{subfigure}{\textwidth}
    \centering
    \includegraphics[width=\textwidth]{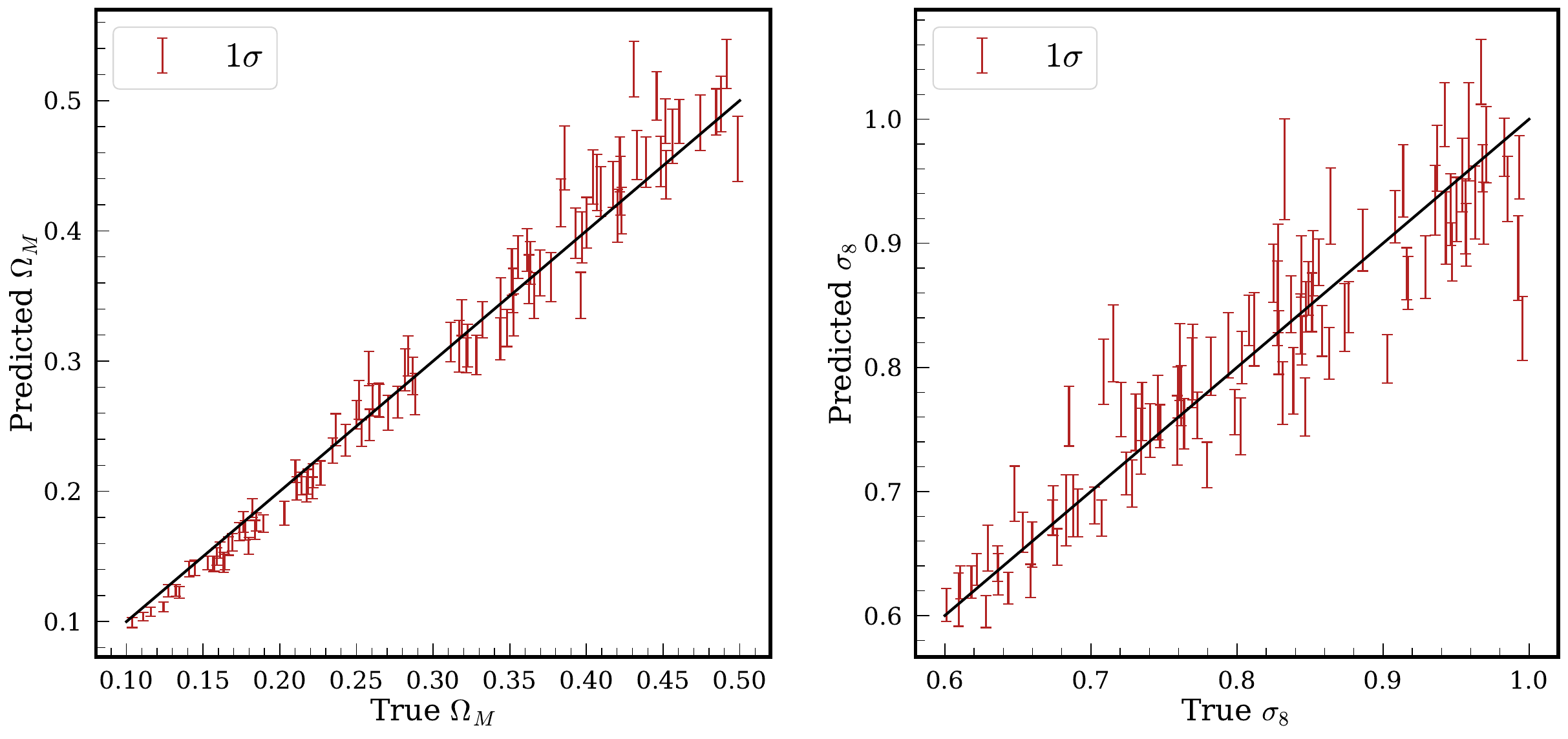}
    \caption{Supervised: trained on IllustrisTNG, tested on SIMBA.}
\end{subfigure}
\begin{subfigure}{\textwidth}
    \centering
    \includegraphics[width=\textwidth]{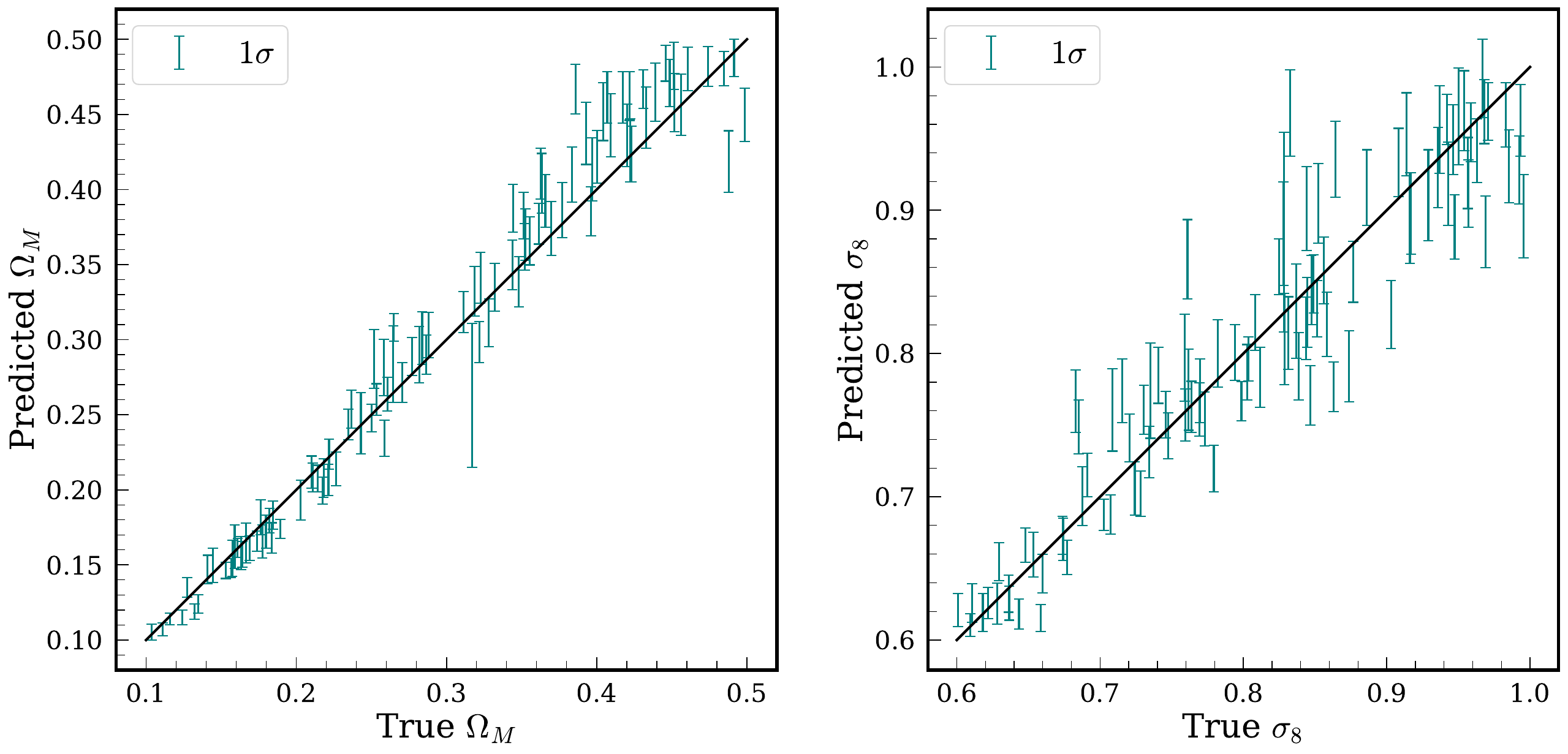}
    \caption{VICReg: trained on IllustrisTNG, tested on SIMBA.}
\end{subfigure}       
\caption{Predicted means and 1-$\sigma$ uncertainties of cosmological parameters $\Omega_M$ and $\sigma_8$ compared to the true values of the parameters for total matter density maps from SIMBA simulations using the models trained on maps from IllustrisTNG simulations. Predictions for the means and variances of the parameters were obtained by training a convolutional neural network in a fully-supervised way (top) and by training simple inference neural network on VICReg summaries (bottom). 
} 
\label{fig:app_train_on_IllustrisTNG_test_on_SIMBA_truth_vs_prediction}
\end{figure*}

\begin{figure*}
\begin{subfigure}{\textwidth}
    \centering
    \includegraphics[width=\textwidth]{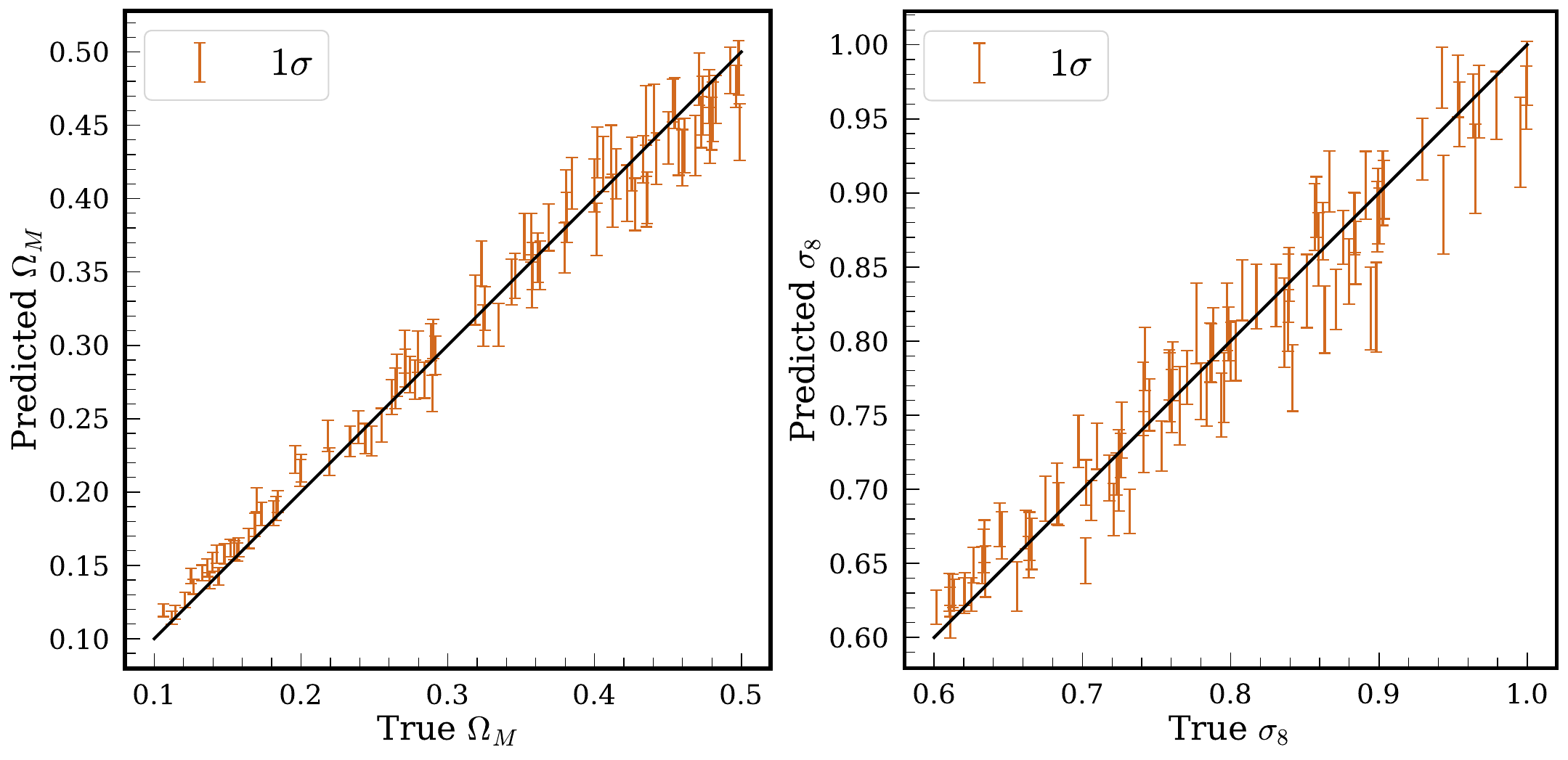}
    \caption{Supervised: trained on SIMBA, tested on IllustrisTNG.}
\end{subfigure}
\begin{subfigure}{\textwidth}
    \centering
    \includegraphics[width=\textwidth]{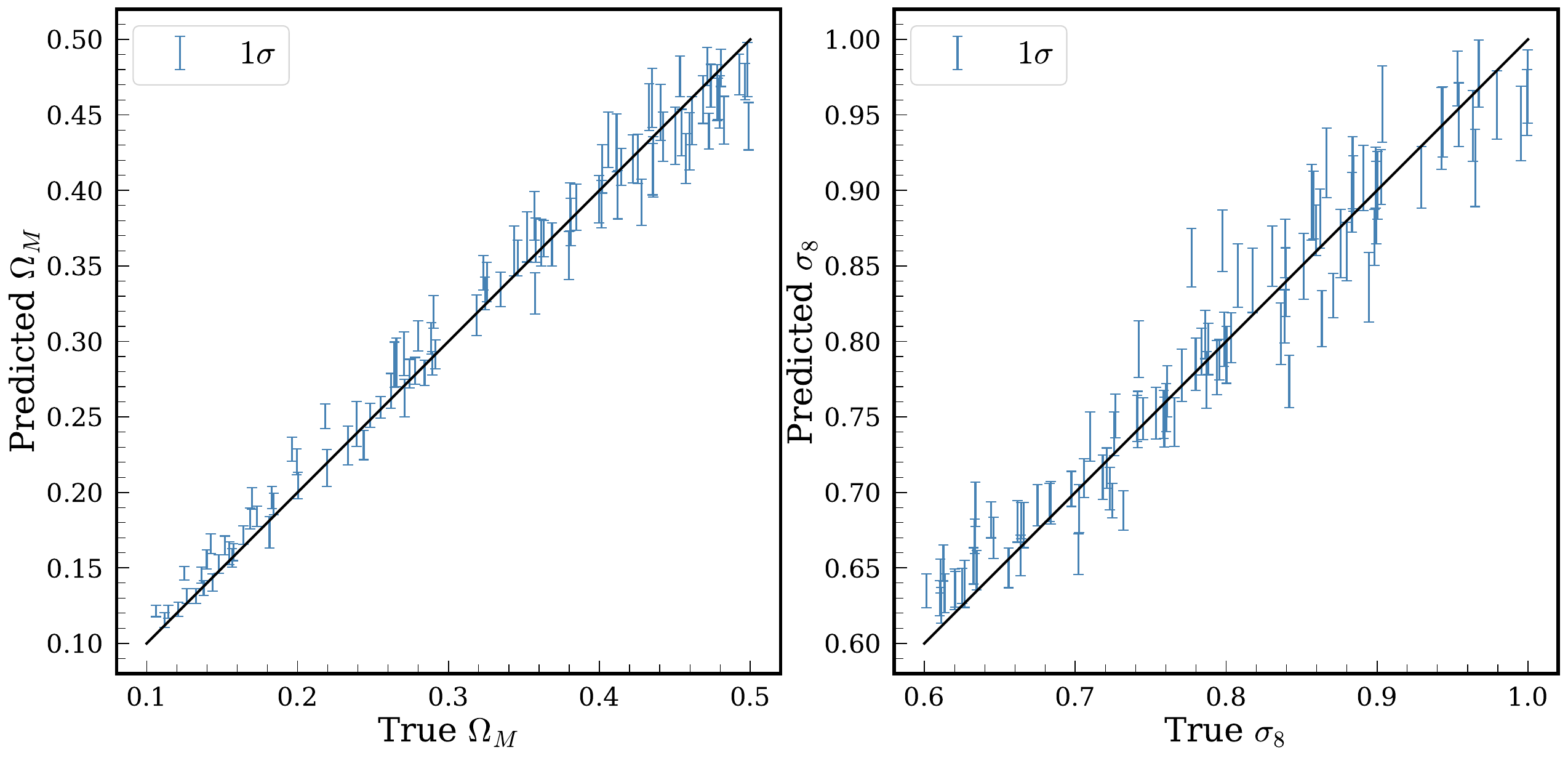}
    \caption{VICReg: trained on SIMBA, tested on IllustrisTNG.}
\end{subfigure}       
\caption{Predicted means and 1-$\sigma$ uncertainties of cosmological parameters $\Omega_M$ and $\sigma_8$ compared to the true values of the parameters for total matter density maps from IllustrisTNG simulations using the models trained on maps from SIMBA simulations. Predictions for the means and variances of the parameters were obtained by training a convolutional neural network in a fully-supervised way (top) and by training simple inference neural network on VICReg summaries (bottom).}
\label{fig:app_train_on_SIMBA_test_on_IllustrisTNG_truth_vs_prediction}
\end{figure*}

We plot the predicted values of $\Omega_M$ and $\sigma_8$ for the two out-of-distribution dataset scenarios for the two methods (VICReg and supervised) on Figures \ref{fig:app_train_on_IllustrisTNG_test_on_SIMBA_truth_vs_prediction} and \ref{fig:app_train_on_SIMBA_test_on_IllustrisTNG_truth_vs_prediction}. 
We notice that while the supervised model predicts the cosmological parameters more accurately overall, the predictions from the inference network show some qualitative similarities with the predictions from the supervised model. 
For instance, on Fig. \ref{fig:app_train_on_IllustrisTNG_test_on_SIMBA_truth_vs_prediction} we find that both models under-predict the value of $\Omega_M$ when $\Omega_M$ is low, and over-predict it when $\Omega_M$ is low. 
Similarly, both models show the reverse trend when predicting $\Omega_M$ on Fig. \ref{fig:app_train_on_SIMBA_test_on_IllustrisTNG_truth_vs_prediction}. This suggests that, with preserving the relevant cosmological information from the maps, the summaries also encode some of the biases present in the simulations they were derived from, since the training of the encoder was not set up in such a way as to create summaries insensitive to the systematic effects present in these simulations.

\begin{table}
\begin{subtable}[h]{\columnwidth}
 \centering
\begin{tabular*}{\textwidth}{c c c c c}
  \hline
  \bf{Method} & \bf{Loss} & \bf{MSE} & \bf{MSE on $\Omega_M$} & \bf{MSE on $\sigma_8$} \\
   & & & (Relative error) & (Relative error) \\ \hline
  VICReg & -2.55 & 4.73 $\times 10^{-4}$ & 3.19 $\times 10^{-4}$  & 6.26 $\times 10^{-4}$ \\
    & & & (4.65\%) & (2.53\%) \\ \hline
  Supervised & -3.29 & 3.92 $\times 10^{-4}$ & 2.55 $\times 10^{-4}$ & 5.28 $\times 10^{-4}$ \\
  & & & (4.36\%) & (2.21\%) \\
  \hline
  \end{tabular*}
 \caption{Trained on SIMBA, tested on IllustrisTNG.}
 \label{tab:app_compare_VICReg_CNN_SIMBA_on_I} 
\end{subtable}

\begin{subtable}[h]{\columnwidth}
\centering
\begin{tabular*}{\textwidth}{c c c c c}
  \hline
  \bf{Method} & \bf{Loss} & \bf{MSE} & \bf{MSE on $\Omega_M$} & \bf{MSE on $\sigma_8$} \\
   & & & (Relative error) & (Relative error) \\ \hline
   VICReg & -2.00 & 9.14 $\times 10^{-4}$ & 5.08 $\times 10^{-4}$  & 13.2 $\times 10^{-4}$ \\
    & & & (5.27\%) & (3.24\%) \\ \hline
  Supervised & -2.54  & 8.06 $\times 10^{-4}$ & 3.92 $\times 10^{-4}$ & 12.2 $\times 10^{-4}$ \\
  & & & (4.96\%) & (3.17\%) \\
  \hline
 \end{tabular*}
 \caption{Trained on IllustrisTNG, tested on SIMBA.}
 \label{tab:app_compare_VICReg_CNN_IllustrisTNG_on_S} 
 \end{subtable}
 \caption{Summary of the performance of the VICReg method and of an equivalent supervised model for inferring cosmological parameters $\Omega_M$ and $\sigma_8$ trained on SIMBA and evaluated on IllustrisTNG and vice-versa.}
\end{table}
\section{Marginalization over Systematics and Nuisance Parameters}\label{sec:BaryonicMarg}

A general task across many fields involves summarization of complex data in a manner that isolates the effects of interesting variations in a model (e.g., parameters of interest) while accounting for the space of variations of uninteresting effects (e.g., nuisance or latent parameters). In the context of parameter inference, this is often accomplished through marginalization (in the Bayesian setting) or profiling (in the frequentist setting). This can be especially challenging if the space of systematic variations is high-dimensional and/or not well-understood.

In cosmological inference, a specific issue involves summarization and inference while accounting for systematic effects, both physical and observational, e.g. the effect of baryonic physics. In recent years, a number of works have investigated robustness of supervised machine learning methods to uncertainties associated with modelling baryonic processes in numerical simulations. 
These processes, which include feedback from Active Galactic Nuclei (AGN) and stellar feedback, affect observables such as the matter power spectrum at nonlinear scales \citep{baryons_chisari2019modelling}. 
Different suites of hydrodynamical simulations take different approaches to modelling baryonic physics effects, which vary in the numerical methods used and their implementation of baryonic effects (or `sub-grid' physics) \citep{IllustisTNG_sims_1,IllustisTNG_sims_2,SIMBA_simulations,baryons_chisari2018impact}. 
As a result, theoretical predictions of cosmological observables from these simulations do not necessarily agree well on small scales, which are most affected by baryonic physics. 
Some studies have found machine learning models that are robust to variations in `sub-grid' physics across different simulations (e.g. \citet{CMD_Mtot_RobustInference,baryonsRobust_villanueva2022inferring,baryonsRobust_desanti2023}). 
Others, however, do not generalize well when applied to data from new, previously unseen suites of simulations (e.g. \citet{baryonsNotRobust_villanueva2022learning,baryonsNotRobust_delgado2023predicting}). 
In order to further address the robustness question, new numerical simulations with distinct `sub-grid' physics models are now being incorporated into the CAMELS simulation suite \citep{baryons_camels_new_suite}. 

The self-supervised method we have introduced offers an avenue to build machine learning models that are insensitive to uncertainties, such as those due to baryonic effects. These methods are designed to compress data into a set of statistics which preserve relevant information and are insensitive to a given set of variations. 
If we are interested in isolating information due to cosmological parameters of interest, then different augmentations used to train an encoder network could be the simulation products from different sets of simulations (such as SIMBA and IllustrisTNG), which share the same cosmological parameters and initial conditions, but follow different `sub-grid' physics prescriptions, or span a range of variations in sub-grid modeling. 
In such a setup, the encoder would learn to produce representations that are insensitive to variations in the implementation of baryonic physics.
Since a large-scale cosmological dataset with the necessary augmentation structure is unavailable at the present time, we motivate this use case with a simple proof-of-principle example in the following subsection.

\subsection{Data}\label{sec:BaryonicMarg_data}

Following \cite{NeuralNetworksOptimalEstimators}, for simplicity and ease of interpretation, we study a toy model of a power spectrum observable represented by a broken power law:
\begin{equation}\label{eq:toy_Pk_model}
P(k) = \begin{cases}
Ak^{B} &\text{$ k \leq k_{\mathrm{pivot}}$}\\
Ck^{D} &\text{$ k > k_{\mathrm{pivot}}$},
\end{cases}
\end{equation}
where $A$ and $B$ are proxies for `cosmological parameters' of interest that describe the amplitude and scale-dependence of the power spectrum on scales that are unaffected by the `baryonic effects'. 
Parameter $D$, on the other hand, is a proxy for the effects of baryons on small scales which, in this toy model, change the slope of the power spectrum. Finally, $C$ is a constant that describes
the amplitude of the power spectrum beyond the pivot scale $k_{\mathrm{pivot}}$. 
It is calculated by requiring that the power spectrum is continuous at the pivot scale: $Ak^{B}_{\mathrm{pivot}} = C k^{D}_{\mathrm{pivot}}$, so the small scales ($k>k_{\mathrm{pivot}}$) carry `cosmological' information about parameters $A$ and $B$ via $C$. 

\begin{figure}
 \includegraphics[width = \columnwidth]{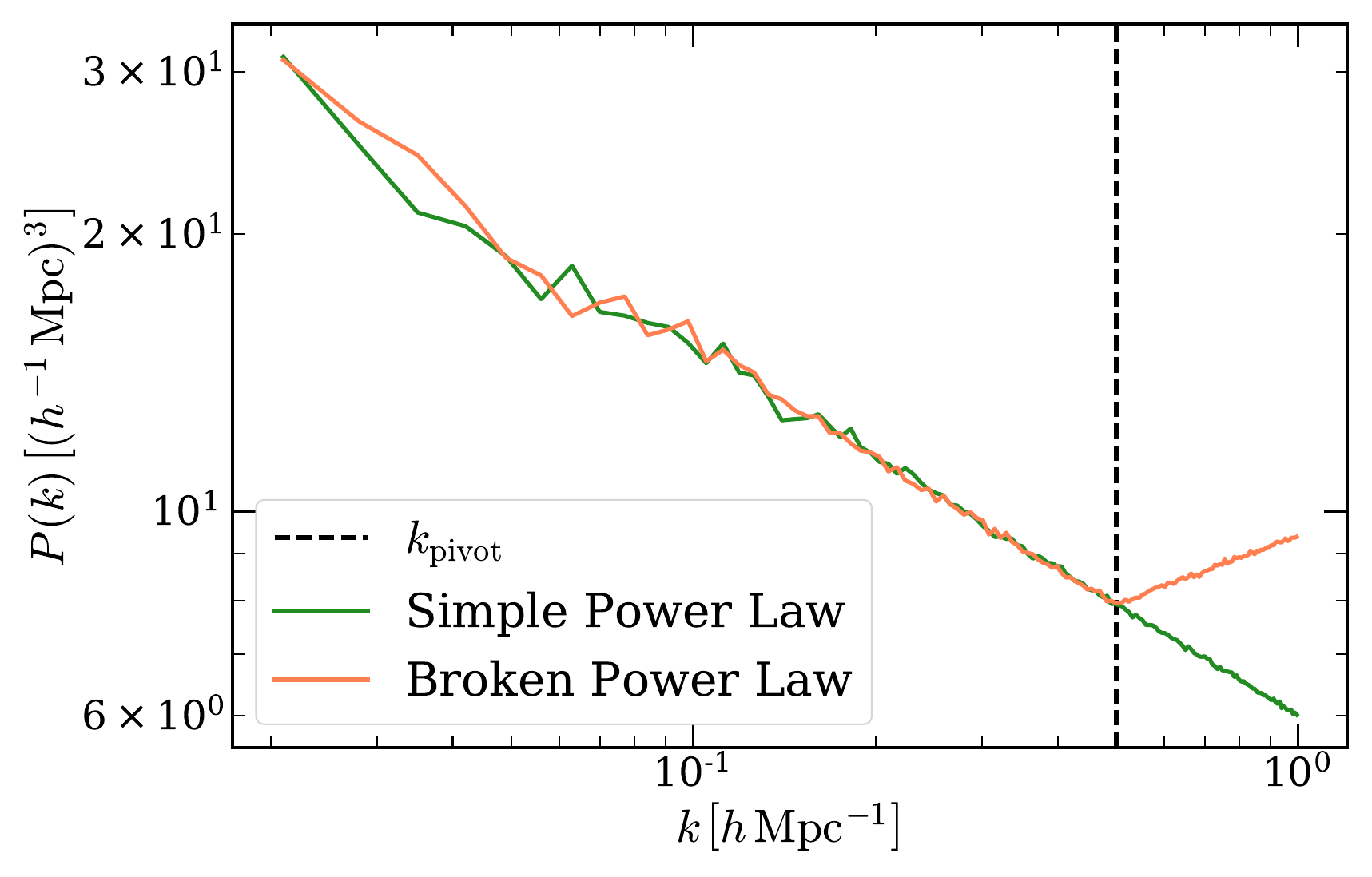}
 \caption{An example of a toy power spectra represented by a simple power law (no `baryonic effects') and a broken power law (`baryonic effects' affect scales past the pivot scale $k_{\mathrm{pivot}}$). The power spectra were generated for the following set of parameters: ${A = 0.6, B = -0.4, D = 0.25}$ (see Eq. \ref{eq:toy_Pk_model}).}
 \label{fig:toy_Pk_example}
\end{figure}

For this simple idealized case, we do not include noise into the model, but add cosmic variance effects. Cosmic variance accounts for the fluctuations or differences from the true values of the power spectrum one would get when measuring the power spectrum $P_{\mathrm{obs}}(k)$ from a realization of a corresponding Gaussian field:
\begin{equation}\label{eq:toy_Pk_cosmic_var}
    P_{\mathrm{obs}}(k) \sim \mathcal{N}\left(P(k), \sigma_k^2\right).
\end{equation}
The variance on the power spectrum is given by $\sigma_k = \sqrt{\frac{2}{N_k}} P(k)$, where $N_k = \frac{4 \pi k^2 k_F}{k_F^3}$ is the number of $k$-modes in a given bin and $k_F$ is the fundamental frequency for a simulation box or a survey. For our dataset, we set $k_F$ to $7\times 10^{-3} h \, \mathrm{Mpc}^{-1}$, $k_{\mathrm{pivot}}$ to 0.5 $h \, \mathrm{Mpc}^{-1}$, and compute the power spectrum for modes in range $k \in [3, 142] k_F$. In Fig. \ref{fig:toy_Pk_example}, we show one example of such power spectrum produced with ${A = 0.6, B = -0.4, D = 0.25}$. On the same figure, for comparison we also show an example of a power spectrum described by a simple power law with the same $A$, $B$ parameters ($D=B$). 

For our dataset, we sample 1,000 different values of the parameters $A$ and $B$ from uniform distributions: $A \in [0.1, 1.0], \, B \in [-1., 0.0]$. 
For each combination of $A$ and $B$, we sample 10 different values of parameter $D$ uniformly at random ($D \in [-0.5, 0.5]$) and generate realizations of the observed power spectra. 
These power spectra, which share $A$ and $B$, but vary $D$, are used as different views when training the VICReg encoder. 
This is the dataset we are most interested in, and we later refer to this case as \textit{`broken power law with varying $D$'}. 

For comparison, we create a second dataset that uses the same 10,000 combinations of parameters $A$ and $B$. 
However, for each combination of $A$, $B$, we sample only one value of parameter $D$ and generate 10 realizations of the corresponding power spectra. 
These realizations, used as different views for training the encoder, share both `cosmological' and `baryonic' parameters, but differ from one another only due to cosmic variance. 
We refer to this case as \textit{`broken power law with constant $D$'}. 

\subsection{VICReg Setup}\label{sec:BaryonicMarg_VICReg}

We use a fully-connected neural network with 7 layers as the encoder network. The encoder compresses the input $P(k)$ vectors of length 140 down to 32-dimensional summaries. The inference network is a simple fully-connected network with 2 layers. We outline the architectures of the two networks in Appendix \ref{appendix:Baryons_NN_Info}.

We use 80\% of the data for training, 10\% for validation, and 10\% for testing. We run the training for the encoder network for 300 epochs with learning rate of $10^{-3}$ and cosine annealing schedule. 
The invariance $\lambda$, variance $\mu$, and covariance $\nu$ weights are set to 15, 15, and 1 respectively in the loss function. The inference network is trained on the summaries for 300 epochs with initial learning rate of $5\times 10^{-4}$, which is multiplied by a factor of 0.3 if the validation loss plateaus for 10 epochs.

\subsection{Results}\label{sec:BaryonicMarg_Results}

We evaluate the overall performance of the VICReg method on the test dataset corresponding to `broken power law with varying $D$' case. 

\begin{figure*}
\centering
 \includegraphics[width =\textwidth]{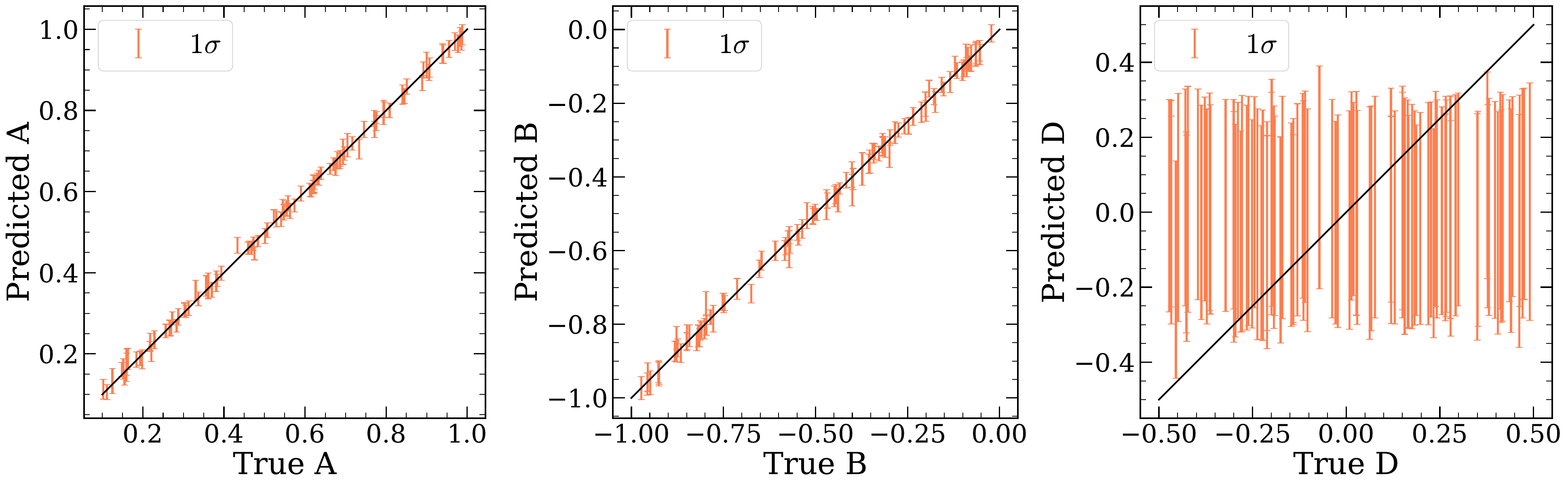}
 \caption{Predicted means and 1-$\sigma$ uncertainties of `cosmological' parameters $A$, $B$ and `baryonic physics' parameter $D$ as a function of the true values of the parameters. Predictions for the means and variances of the parameters are plotted for the dataset which is referred to as the `broken power law with varying $D$' in this Section.}
 \label{fig:toy_Pk_3params_inference}
\end{figure*}

We begin by analyzing the results of cosmological parameter inference from the summaries. We plot the predicted values of parameters $A, B, \mathrm{and} \, D$ against the true values of these parameters in Fig. \ref{fig:toy_Pk_3params_inference}. 
We find that the regression network trained on the summaries is able to predict the `cosmological parameters', with relative errors on $A$ and $B$ $2.1\%$ and $3.9\%$ respectively. 
However, the network is not able to infer any information regarding the `baryonic effects' parameter $D$ from the summaries. 
This is promising, since the augmentations were specifically designed so that the representations produced by the VICReg encoder network would be insensitive to variations due to `baryonic effects' while retaining information about the relevant `cosmological parameters'. 

Since small scales ($k > k_{\mathrm{pivot}} $) still contain information about the `cosmological' parameters, a potential worry about our prescription is that it blindly ignores the small scales in the data and the information they contain, learning a trivial prescription to ignore `baryonic' effects. Instead, we would like the method to learn to disentangle the `cosmological' information from the `baryonic' information on these scales. 
We study how much, if at all, the representations produced via self-supervision depend on the power spectrum at different scales $k$ using two different metrics of statistical dependence: distance correlation and mutual information. We will see that the learned summaries retain correlations with the input power spectra at input scales beyond the pivot scale, showing the retention of `cosmological' information despite having no sensitivity to `baryonic' parameters.

\subsubsection{Distance Correlation}\label{sec:BaryonicMarg_dcor}

Distance correlation is a statistical measure of dependence between two random variables that captures both linear and non-linear associations between the variables \citep{distanceCorrelation_szekely2007}. For a pair of random variables $X$ and $Y$, the distance correlation $\mathcal{R}(X, Y)$ can be computed as follows:

\begin{equation*}\label{eq:toy_Pk_distance_corr}
\mathcal{R}(X, Y) = \begin{cases}
\frac{\mathcal{V}^2(X, Y)}{\sqrt{\mathcal{V}^2(X)} \sqrt{\mathcal{V}^2(Y)}}, & \mathcal{V}^2(X) \mathcal{V}^2(Y) > 0\\
0, & \mathcal{V}^2(X) \mathcal{V}^2(Y) = 0,
\end{cases}
\end{equation*}

where $\mathcal{V}^2(X, Y)$ is the (squared) distance covariance between $X$ and $Y$, and $\mathcal{V}^{2}(X)$, $\mathcal{V}^{2}(Y)$ are the distance variances of $X$, $Y$. 
Empirically, the distance covariance for a statistical sample of N pairs of random vectors $(X_k, Y_k), \, k = 1, 2, .., N$ can be computed as an average of the so-called doubly-centered distances $A_{k l}, B_{k l}$:
\begin{equation}\label{eq:distance_cov_definition}
\mathcal{V}_N^2(X, Y)=\frac{1}{N^2} \sum_{k, l=1}^n A_{k l} B_{k l},
\end{equation}
where the doubly-centered distances $A_{k l}, B_{k l}$ are defined as: 
\begin{equation}\label{eq:distance_cov_doubly_cent_dists_definition}
A_{j, k}=a_{j, k}-\bar{a}_{j .}-\bar{a}_{\cdot k}+\bar{a}_{. .}, \quad B_{j, k}=b_{j, k}-\bar{b}_{j .}-\bar{b}_{. k}+\bar{b}_{. .},
\end{equation}
and $a_{j, k}, b_{j, k}$ are matrices containing pairwise Euclidean distances between the vectors:
\begin{equation}\label{eq:distance_cov_distance_matrix_definition}
\begin{aligned}
a_{j, k} & =\left\|X_j-X_k\right\|, \, \, b_{j, k}=\left\|Y_j-Y_k\right\|,
\end{aligned}
\end{equation} 
and $\bar{a}_{j .}$ and $\bar{b}_{j .}$ are the row means, $\bar{a}_{\cdot k}$ and $\bar{b}_{\cdot k}$ are the column means, $\bar{a}_{. .}$ and $\bar{b}_{. .}$ are the overall means of the matrix distances $a_{j, k}, b_{j, k}$. 

We refer the reader to \cite{distanceCorrelation_szekely2007} for the full definitions and properties of these quantities. In general, the distance correlation $\mathcal{R}(X, Y)$ can vary between 0 and 1, with $\mathcal{R}(X, Y) = 0$ corresponding to the case of two independent variables and $\mathcal{R}(X, Y) = 1$ indicating linear relationship between the variables. 

\begin{figure}
\centering
 \includegraphics[width =\columnwidth]{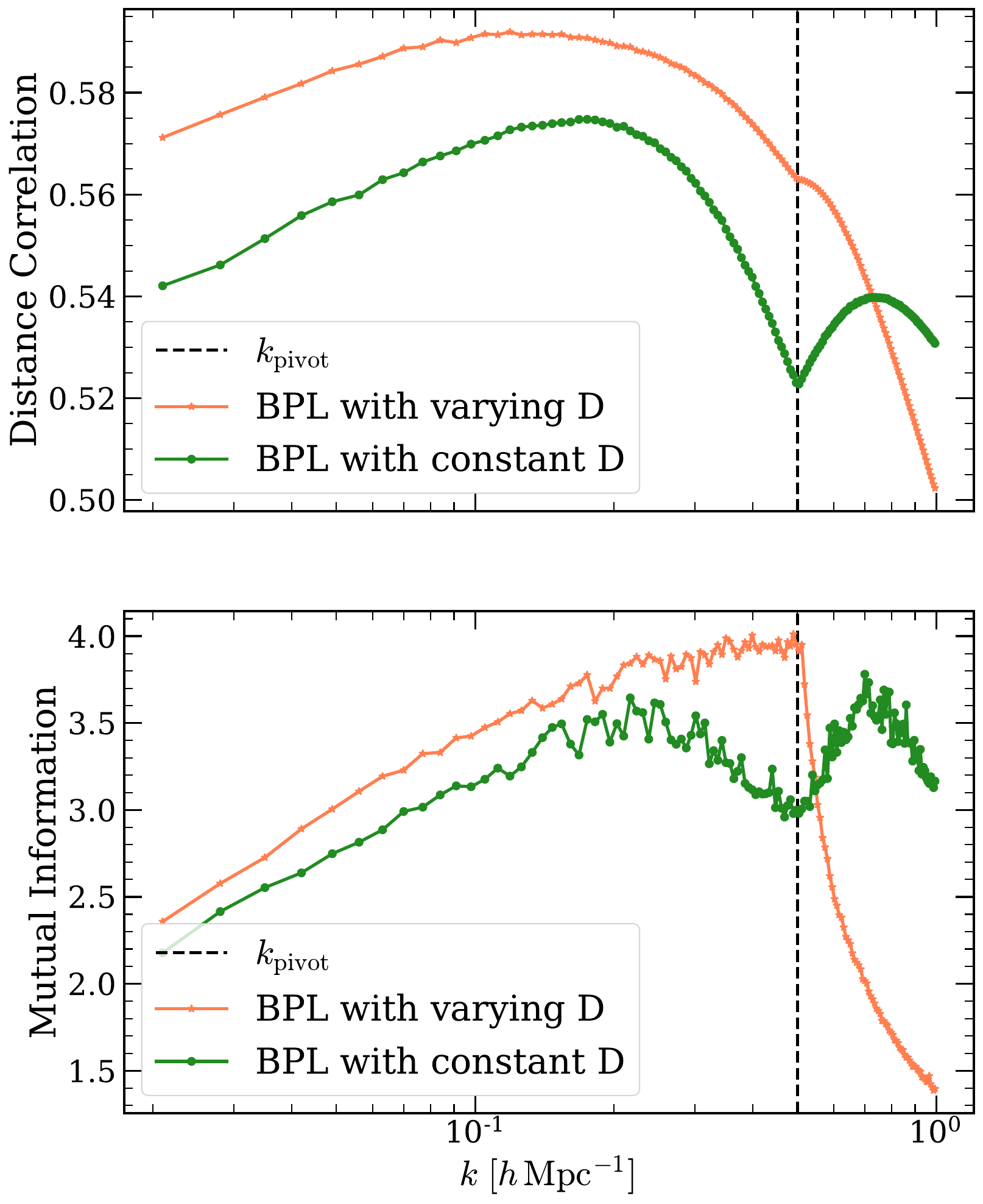}
 \caption{Comparison of the distance correlation (top) and of the estimated mutual information (bottom) between the VICReg summaries and power spectra values in different $k$-bins for the two datasets: `broken power law (BPL) with varying $D$' and `broken power law (BPL) with constant $D$'.}
 \label{fig:toy_Pk_metrics}
\end{figure}

Figure \ref{fig:toy_Pk_metrics} shows the distance correlation between the VICReg summaries and the power spectra values in a given $k$-bin. Here we consider summaries from the two cases outlined in Section \ref{sec:BaryonicMarg_data}: `broken power law with varying $D$' and `broken power law with constant $D$'. The distance correlation between the summaries and different $k$-modes follows similar behavior for the two datasets prior to the pivot scale. 
Past $k_{\mathrm{pivot}}$, the two curves start to differ: distance correlation for the summaries trained on the power spectra with varying $D$ declines rapidly, whereas, for the summaries trained on the power spectra with constant $D$, the distance correlation starts to increase again. 

This behaviour is consistent with the expected results, given the two different VICReg setups used to obtain the summaries. On large scales $k < k_{\mathrm{pivot}}$ both types of power spectra carry relevant information about `cosmological parameters', since at these scales the power spectra depend only on $A$ and $B$. The small scales, however, contain information about both `cosmological' and `baryonic' parameters. 
In the first case, the `baryonic' information is considered irrelevant, so the summaries should depend less on power spectra values past $k_{\mathrm{pivot}}$. For the second case, the same `baryonic' information is considered relevant since it is present across different augmentations of the same object, hence the increase in the value of distance correlation between the summaries and the power spectra values on small scales.

\subsubsection{Mutual Information}\label{sec:BaryonicMarg_MI}

Mutual information (MI) is another measure of non-linear dependence between random variables. It measures the amount of information (in natural units or nats) one gains about one random variable by `observing' the other random variable. Similar to the distance correlation, mutual information extends beyond linear correlations, captures the full dependence between the two variables, and is zero only if the two variables are independent. 
For a pair of variables $X$ and $Y$, mutual information $I(X, Y)$ is defined as:
\begin{equation}
\begin{split}
    I(X, Y) & \equiv D_\mathrm{KL}\left(P_{XY} || P_{X} \otimes P_Y\right) \\
    & = \int{ P_{XY}(x,y) \log \frac{P_{XY}(x, y)}{P_X (x) P_Y (y)} dx dy},
\end{split}
\end{equation}
i.e., the Kullback-Leibler divergence $D_\mathrm{KL}$ between the joint distribution $P_{XY}$ and the product of marginals $P_X$ and $P_Y$. 
The integral is taken over the joint support of $X$ and $Y$.
For a comprehensive review of mutual information and its properties, we refer the reader to, for instance, \cite{mutualInfo_vergara2014review}. 

Mutual information has been extensively leveraged in astrophysics and cosmology (e.g. \citet{MI_example_Pandey}, \citet{MI_example_Jeffrey}, \citet{MI_example_Malz}, \citet{MI_example_Upham}, \citet{MI_example_LucieSmith}, \citet{MI_example_sui2023evaluating}). However, estimating mutual information between high-dimensional random variables is challenging \citep{mutualInfo_estimation_Paninski}. 
Several estimators of mutual information have been proposed to address this (e.g. \citet{MI_methods_kraskov}, \citet{MINE}, \citet{MI_methods_holmes}, \citet{MI_methods_piras}).

We use the estimator MINE (Mutual Information Neural Estimation) \citep{MINE} to compute a mutual information estimate between the VICReg summaries and power spectra in different $k$-bins. MINE trains a neural surrogate to distinguish between samples from the joint distribution and the independent marginal distributions to maximize a lower bound on the mutual information; see \citet{MINE} for more information. 

In Fig. \ref{fig:toy_Pk_metrics}, we plot the estimated mutual information between the VICReg summaries and the values of power spectra values in different $k$-bins. Similarly to Sec. \ref{sec:BaryonicMarg_dcor}, we consider two cases: `broken power law with varying $D$' and `broken power law with constant $D$'. 
We notice that the mutual information and distance correlation follow a similar pattern. 
Past the pivot scale $k_\mathrm{pivot}$, mutual information estimate decreases in magnitude in the first case, while in the second case there is also a slight enhancement of MI on small scales, which mirrors the enhancement in the distance correlation. 
The suppression of mutual information on small scales in the first case aligns with what one might expect: VICReg training was set up to generate summaries insensitive to `baryonic physics', which only affects $k > k_{\mathrm{pivot}}$, so the summaries should show less dependence (lower MI) on small scales. We note that in the case with constant $D$ (i.e., when summaries are expected to carry information about all parameters of interest), for both MI and distance correlation, the correlation measure peaks at an intermediate scale in the $k$-range where it is expected to be sensitive to specific parameters ($k > k_\mathrm{pivot}$ for the `baryonic' parameter, and $k < k_\mathrm{pivot}$ for the `cosmological' parameter).

In summary, Figs. \ref{fig:toy_Pk_3params_inference} and \ref{fig:toy_Pk_metrics} show that the summaries learned through self-supervision preserve relevant `cosmological' information, while being insensitive to the variations in `baryonic effects' by disentangling the signal coming from `cosmological' parameters from the information about `baryonic' parameters, instead of entirely ignoring the small scales.
While in this section, for the ease of interpretation, we considered a simple toy model as our data vector, it would be interesting to investigate further if the same holds for more complex and realistic data vectors, such as cosmological power spectra with various non-linear corrections due to baryonic effects or matter density maps from hydrodynamical simulations with different baryonic physics prescriptions.

\subsubsection{Comparison to the Supervised Baseline Model}
For completeness, we also compare the performance of the VICReg method and of supervised learning and find that, similarly to the results from Section \ref{sec:VICreg_data_compression}, an equivalent supervised model slightly outperforms the self-supervised model. In particular, when marginalizing over the `baryonic physics' parameter $D$ and inferring only the `cosmological' parameters, the relative errors on $A$ and $B$ from the supervised model are $1.30\%$ and $2.98\%$ respectively, while the relative errors from the self-supervised model are at the level of $1.90\%$ and $3.39\%$ respectively. 
\cite{NeuralNetworksOptimalEstimators} found that for the toy $P(k)$ model studied in this Section, neural networks trained in a supervised manner are able to marginalize out the scales which are impacted by baryonic effects and to obtain relevant cosmological information even from the scales strongly affected by these effects. While we cannot directly compare our results to \cite{NeuralNetworksOptimalEstimators} due to differences in the datasets and the neural network architectures used in the study, we find that supervised baseline models place tight and accurate constraints on the `cosmological' parameters when marginalizing over `baryonic' effects, which is consistent with their findings.

\section{Cosmological Parameter Inference with Sequential Simulation-Based Inference} \label{sec:application_to_SBI}

We have so far focused on cases where compressed summaries are used to build \emph{amortized} estimators for the parameter posterior distribution, i.e. ones that can be deployed for inference on any new data point. Many cosmological applications can benefit instead from \emph{targeted} inference -- learning an estimator specific to a given observation. This can be challenging when one only has access to a fixed set of simulations. We show here how our compression scheme can be used to build a generative model for use as an emulator of summaries, to be then used for targeted inference using \emph{sequential} simulation-based inference. 

Simulation-based inference (SBI) refers to a broad set of methods that are designed to infer parameters of interest $\Vec{\theta}$ when the likelihood describing the observed data $\Vec{x}_o$ is unknown or intractable. SBI techniques rely on forward models or \textit{simulators} which probabilistically model the data generation process and thus implicitly define the likelihood $p(\Vec{x}_o|\Vec{\theta})$ of the data given a set of input parameters. By aggregating and analyzing the samples generated from the simulator, SBI techniques approximate the posterior distribution of the input parameters $\Vec{\theta}$. In the recent years, advances in machine learning have led to development of new neural SBI methods that address the shortcomings associated with traditional SBI methods, such as Approximate Bayesian Computation (ABC), and enable efficient and accurate posterior inference, even for complex high-dimensional distributions. We refer the reader to \citet{sbi_frontierSBI_cranmer2019} for a recent review of neural SBI and associated developments. 

Here, we use the Sequential Neural Posterior Estimation (SNPE) method \citep{sbi_SNPE_papamakarios2016fast, sbi_SNPE_lueckmann2017flexible, sbi_SNPE_greenberg2019automatic} to perform parameter inference. SNPE sequentially (in rounds) generates $N$ parameter-data pairs $\{\Vec{\theta}_i, \Vec{x}_i\}_{i=1}^{N}$ with the simulator and uses a neural network, such as a normalizing flow \citep{sbi_normalizingFlows_papamakarios2021}, to approximate the true posterior distribution $p(\Vec{\theta}|\Vec{x}_o)$. While the first batch of the parameter-data pairs are drawn from the prior, in subsequent inference rounds the SNPE algorithm draws new samples from a modified proposal distribution obtained by conditioning the approximate posterior estimator on the observation of interest $\Vec{x}_o$. This technique, which is an instance of active learning, allows the neural network to learn from the more informative samples, which are more likely to produce data vectors $x$ similar to the observed data $\Vec{x}_o$, reducing the total number of simulations required to accurately approximate the true posterior. We use the default implementation of the SNPE algorithm provided by the \texttt{sbi} package \citep{sbi_package_tejero-cantero2020sbi}. 

While strategies such as `active learning' can significantly speed up the inference process, one potential bottleneck for an SBI-based pipeline is the computational complexity of the simulator. Although the mock data, such as lognormal overdensity fields, can be generated fairly quickly, if one were to perform SBI on more realistic datasets, such as density fields from hydrodynamical or N-body simulations, obtaining a single sample from the simulator would take a non-trivial amount of time and computational resources. We address this problem by training an emulator using a normalizing flow \citep{sbi_normalizingFlows_papamakarios2021,rezende2015variational} to model the distribution of the summaries $\Vec{S}$ given a set of parameters $\Vec{\theta}$. The advantage of training the emulator on the summaries $S$ as opposed to the input maps themselves is that the summaries contain most (or ideally all) of the information from the maps, while, due to their lower-dimensionality, their distribution should be easier to estimate accurately than that of full maps. Once trained, the emulator can produce the summaries on-the-fly, which addresses the computational bottleneck. 

\begin{figure}
 \includegraphics[width=\columnwidth]{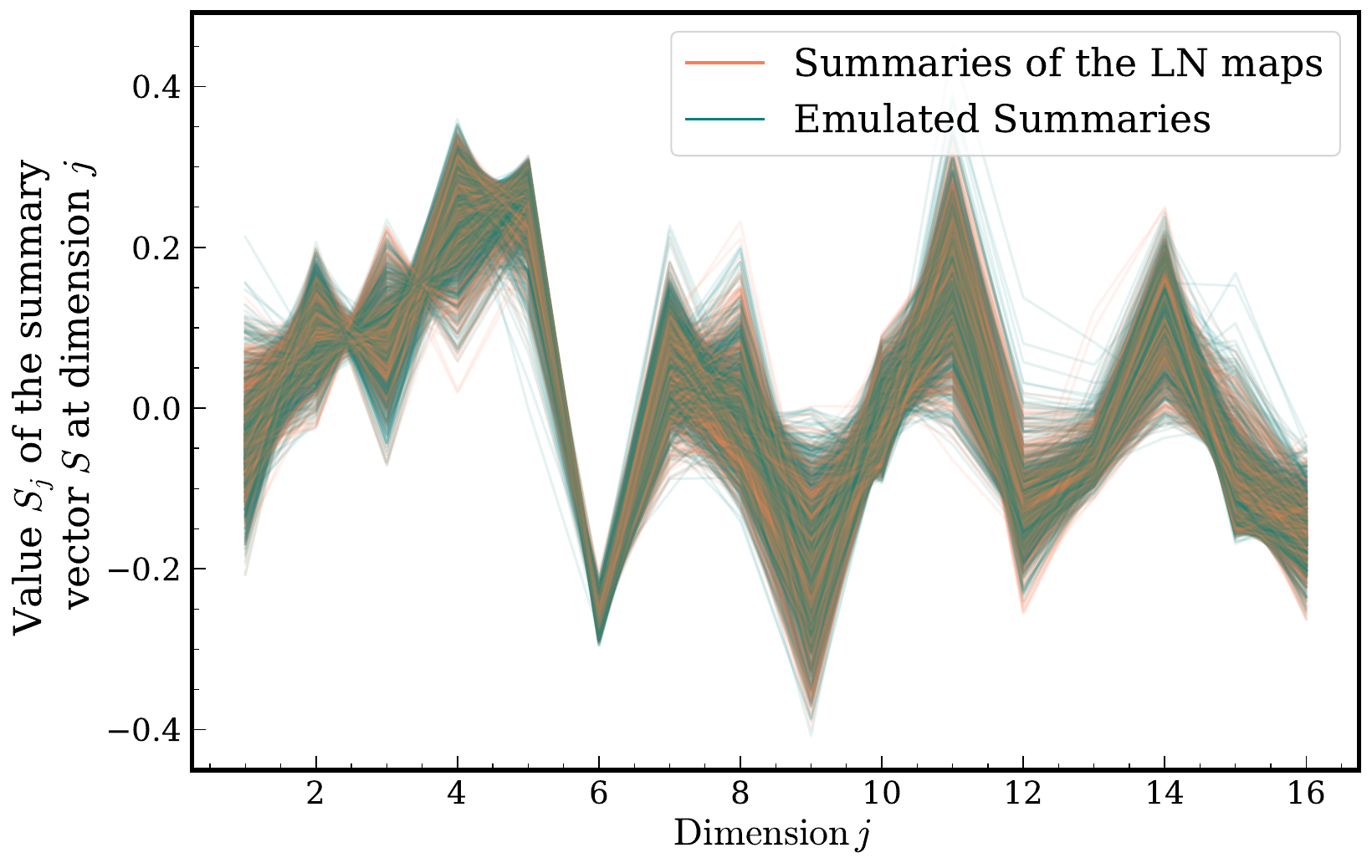}
 \caption{Comparison of the summaries from the trained emulator (teal) and the actual summaries (orange) of lognormal maps (LN) with $\Omega_M = 0.3$ and $\sigma_8 = 0.8$.}
 \label{fig:LN_fields_emulated_summaries}
\end{figure}

We construct the emulator from a stack of 8 masked autoregressive flows \citep{sbi_masked_flows_papamakarios2017}. The emulator is trained on the VICReg summaries of the lognormal maps from the dataset described in Sec. \ref{sec:LognormalFields_Data}. We train the emulator for 300 epochs using \texttt{AdamW} optimizer, with learning rate of $2 \times 10^{-3}$ and cosine annealing schedule. We compare samples of summaries from the trained emulator to the summaries computed on lognormal maps in Fig. \ref{fig:LN_fields_emulated_summaries} for input parameters $\Vec{\theta}_o = \{\Omega_M = 0.3, \, \sigma_8=0.8\}$. By inspection, we find that the emulator can successfully reproduce key features of the distribution of the actual summaries.

\begin{figure}
 \includegraphics[width=\columnwidth]{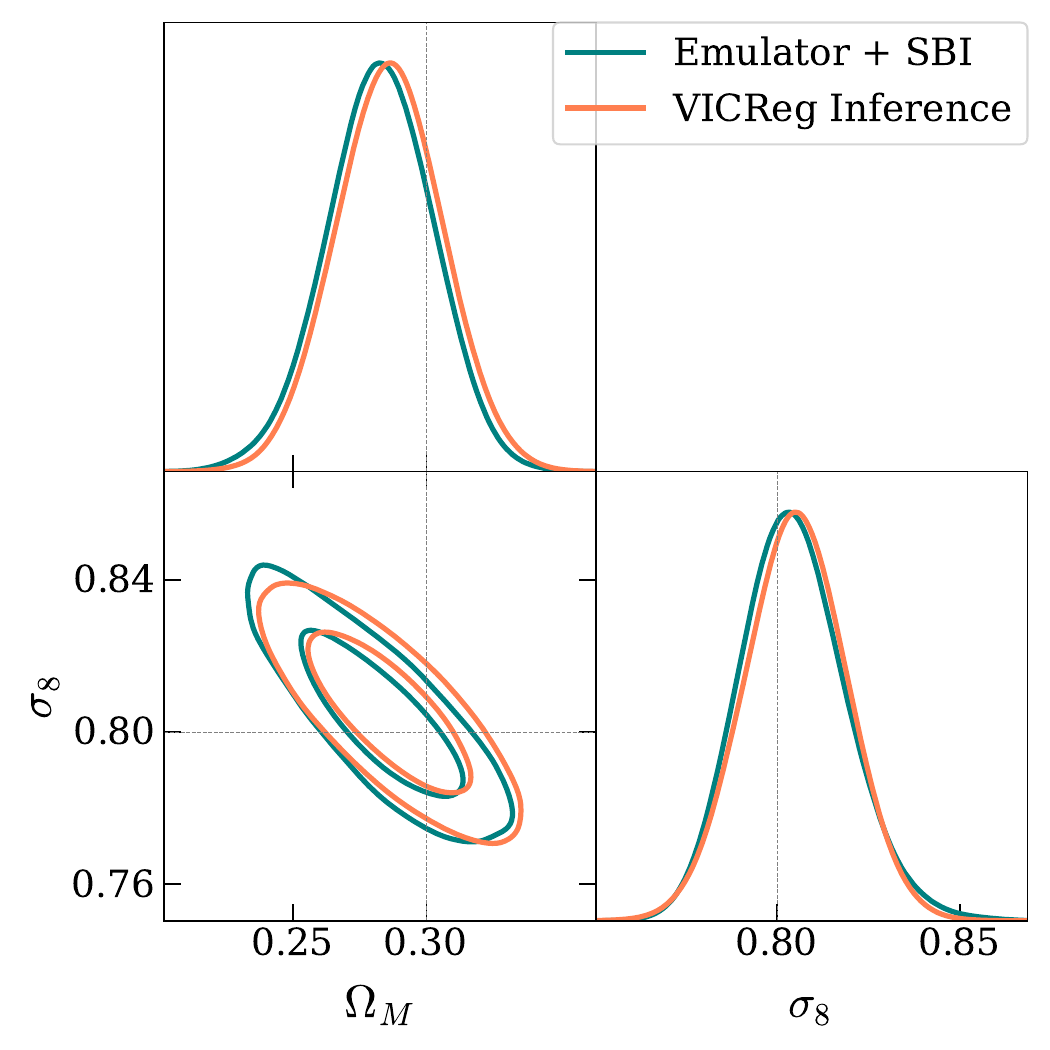}
 \caption{Posterior probability distributions of the input parameters $\theta$, inferred with the Sequential Neural Posterior Estimation algorithm and a trained emulator (teal) and with an inference network from Sec. \ref{sec:LognormalFields_Results} (orange). The results shown on the plot were obtained for a lognormal map generated with $\Omega_M = 0.3$ and $\sigma_8 = 0.8$.}
 \label{fig:LN_fields_posterior_emulator}
\end{figure}

We use the trained emulator as the simulator for the SBI pipeline. Our observed data vector $\Vec{x}_o$ is a summary of a randomly generated lognormal overdensity map with $\Vec{\theta}_o = \{\Omega_M = 0.3, \, \sigma_8=0.8\}$. We run the SNPE algorithm for 10 rounds, drawing 1000 parameter-data pairs in each round and using these simulated pairs to estimate the posterior distribution of the input parameters $\Vec{\theta}$ at the end of each round of inference. Figure \ref{fig:LN_fields_posterior_emulator} shows the approximated posterior inferred with the SNPE algorithm (teal) compared to that obtained using the inference network from Sec. \ref{sec:LognormalFields_Results} (orange). We check that both posteriors are well-calibrated using simulation-based coverage calibration (SBCC) procedure \citep{sbi_deistler2022truncated} in Appendix \ref{appendix:posterior_calib}. The true input parameters $\Vec{\theta}_o$ are seen to lie well within the posterior contours, and the constraints obtained using the inference network are consistent with the SNPE-informed constraints. 
While this is a strictly proof-of-concept application of the VICReg-constructed summaries within the SBI framework, it demonstrates potential utility of the summaries for the downstream task of targeted posterior density estimation.

\section{Discussion and Conclusion}\label{sec:Conclusions}

We have introduced a self-supervised machine learning approach for general-purpose representation learning from simulated data, with a focus on data compression as an application. 
The method uses simulation-based augmentations in order to learn representations based on inherent symmetries of the data (i.e., data variations that correspond to the same underlying physics of interest). 
Applied in the context of cosmology, we have shown the potential of the method for massive compression of cosmological simulations into a set of sufficient summary representations as well as their use for downstream parameter inference. In addition, we have shown how the method can be used to produce representations that are insensitive to a set of systematic variations or nuisance parameters.
The learned summaries can be used for a variety of downstream tasks in cosmology and astrophysics, such as classification, source identification and characterization, anomaly or outlier detection, and parameter inference, as well as in conjunction with other machine learning methods such as simulation-based inference.
In this work, however, we focused on using the compressed summaries specifically for the downstream parameter inference.

Using VICReg  as the backbone method, we showed that summaries learned via self-supervision can be used to faithfully reproduce parameter constraints expected via the Fisher information (in the case of lognormal fields data). 
Considering the total matter density maps from hydrodynamic simulations in the CAMELS project \citep{CMD,CAMELS}, we found that, even when applied to this more complex dataset with fewer data samples, our method is able to construct informative summaries that achieve parameter inference performance on par with a fully-supervised baseline. 

Recent works \citep{CMD_Mtot_RobustInference,Robustness_wadekar2022_SZ_scatter,Robustness_shao2023_Halos,baryonsNotRobust_delgado2023predicting} have explored the robustness of supervised methods to variations in subgrid physics models by applying the neural networks trained on observables from one simulation suite to a dataset from a different simulation suite containing separate models of sub-grid physics. 
Here, we do not make such robustness comparisons, since the self-supervision setup we used was designed with the idea of learning summaries of maps that are invariant to random spatial fluctuations and projections rather than variations in subgrid physics models. 
However, it would be interesting to apply this method to create summaries that are robust to differences in the subgrid physics. In this case, the different augmentations used to train the encoder could correspond to the maps from simulations that share the same cosmological parameters but are run using different codes that differ, for instance, in their implementation of baryonic effects due to stellar and AGN feedback. 

Through an illustrative proof-of-principle example, we finally explored the potential of our method to produce summaries that are robust to certain nuisance parameters or systematic uncertainties, finding that the method is able to produce robust representations that can effectively disentangle the effects of nuisance parameters from those corresponding to parameters of interest.
These results suggest that self-supervision can be used to reduce the sensitivity of machine learning models to physical and observational systematic uncertainties, such as those due to baryonic physics. A more comprehensive study of this potential is warranted and is contingent on the availability of cosmological simulations that include the necessary diversity of variations.

While additional follow-up studies are necessary before deploying self-supervised learning methods on real cosmological data, with the influx of new survey data and simulations products, as well rapid advances in novel machine learning techniques, self-supervised learning methods have the potential to accelerate a wide variety of tasks in cosmology via learning useful representations.

\section*{Acknowledgements}

This work was partially supported by the National Science Foundation under Cooperative Agreement PHY-2019786 (The NSF AI Institute for Artificial Intelligence and Fundamental Interactions, \url{http://iaifi.org/}).
This material is based upon work supported by the U.S. Department of Energy, Office of Science, Office of High Energy Physics of U.S. Department of Energy under grant Contract Number DE-SC0012567. The computations in this paper were run on the FASRC Cannon cluster supported by the FAS Division of Science Research Computing Group at Harvard University.

\section*{Data Availability}
The code used to reproduce the results of this paper is available at \url{https://github.com/AizhanaAkhmet/data-compression-inference-in-cosmology-with-SSL}. 
This research made extensive use of the \texttt{Jupyter}~\citep{Kluyver2016JupyterN},  \texttt{Matplotlib}~\citep{Hunter:2007}, 
\texttt{Numpy}~\citep{harris2020array}, 
\texttt{scikit-learn}~\citep{scikit-learn}, 
\texttt{powerbox}~\citep{Murray2018_software_powerbox},
\texttt{pyccl}\footnote{\url{https://github.com/LSSTDESC/CCL}}~\citep{Chisari_software_ccl}, 
\texttt{Pylians}~\citep{Pylians}, 
\texttt{sbi}\footnote{\url{https://www.mackelab.org/sbi/}}~\citep{sbi_package_tejero-cantero2020sbi}, 
\texttt{PyTorch}~\citep{PytorchDLL}, 
\texttt{PyTorch-Lightning}~\citep{william_falcon_2020_3828935}, 
and \texttt{Scipy}~\citep{2020SciPy-NMeth} packages.

\bibliographystyle{mnras}
\bibliography{refs} 

\appendix

\section{Neural Network Architectures for studying Lognormal Density Maps}\label{appendix:LN_fields_NN_Info}

\subsection{Encoder Network}
Here we present the detailed architecture of our model, which combines a set of convolutional layers with a few fully connected layers at the end of the network. 
We denote the convolutional layers by `CN (kernel size, stride, padding)' and fully-connected layers by `FC (input size, output size)'. The final output of the network has size $\mathrm{n}_\mathrm{out}$.  \\ \\
1. Input: $C \times 100 \times 100 \rightarrow$ \\
2. $\mathrm{CN}\, (3,1,1) \rightarrow \mathrm{H} \times 100 \times 100 \rightarrow$ \\
3. BatchNorm2d $\rightarrow$ LeakyReLU $\rightarrow$ \\
4. $\mathrm{CN}\, (3,1,1) \rightarrow \mathrm{H} \times 100 \times 100 \rightarrow$ \\
5. BatchNorm2d $\rightarrow$ LeakyReLU $\rightarrow$ \\ 
6. AvgPool2d (2, 2) $\rightarrow \mathrm{H} \times 50 \times 50 \rightarrow$ \\ 
7. $\mathrm{CN}\, (3,1,1) \rightarrow \mathrm{2H} \times 50 \times 50 \rightarrow$ \\
8. BatchNorm2d $\rightarrow$ LeakyReLU $\rightarrow$ \\
9. $\mathrm{CN}\, (3,1,1) \rightarrow \mathrm{2H} \times 50 \times 50 \rightarrow$ \\
10. BatchNorm2d $\rightarrow$ LeakyReLU $\rightarrow$ \\
11. AvgPool2d (2, 2) $\rightarrow \mathrm{H} \times 25 \times 25 \rightarrow$ \\
12. $\mathrm{CN}\, (2,1,1) \rightarrow \mathrm{4H} \times 24 \times 24 \rightarrow$ \\
13. BatchNorm2d $\rightarrow$ LeakyReLU $\rightarrow$ \\
14. $\mathrm{CN}\, (3,1,1) \rightarrow \mathrm{4H} \times 24 \times 24 \rightarrow$ \\
15. BatchNorm2d $\rightarrow$ LeakyReLU $\rightarrow$ \\
16. AvgPool2d (2, 2) $\rightarrow \mathrm{4H} \times 12 \times 12 \rightarrow$ \\
17. $\mathrm{CN}\, (3,1,0) \rightarrow \mathrm{8H} \times 10 \times 10 \rightarrow$ \\
18. BatchNorm2d $\rightarrow$ LeakyReLU $\rightarrow$ \\
19. $\mathrm{CN}\, (3,1,0) \rightarrow \mathrm{8H} \times 8 \times 8 \rightarrow$ \\
20. BatchNorm2d $\rightarrow$ LeakyReLU $\rightarrow$ \\
21. AvgPool2d (2, 2) $\rightarrow \mathrm{8H} \times 4 \times 4 \rightarrow$ \\
22. $\mathrm{CN}\, (3,1,0) \rightarrow \mathrm{8H} \times 2 \times 2 \rightarrow$ \\
23. BatchNorm2d $\rightarrow$ LeakyReLU $\rightarrow$ \\
24. AvgPool2d (2, 2) $\rightarrow \mathrm{16H} \times 1 \times 1 \rightarrow$ \\
25. Flatten tensor $\rightarrow$ LeakyReLU $\rightarrow$ \\
26. FC (16H, 16H) $\rightarrow$  LeakyReLU $\rightarrow$ \\
27. FC (16H, 16H) $\rightarrow$  LeakyReLU $\rightarrow$ \\
28. FC (16H, 16H) $\rightarrow$  LeakyReLU $\rightarrow$ \\
29. FC (16H, 16H) $\rightarrow \mathrm{n}_{\mathrm{out}}$

\subsection{Inference Network}
1. $n_{\mathrm{out}}$ $\rightarrow$ FC($n_{\mathrm{out}}$, 16 $n_{\mathrm{out}}) \, \rightarrow$ BatchNorm $\rightarrow$ \\
2. FC($16 \, n_{\mathrm{out}}, \, 16 \, n_{\mathrm{out}} ) \rightarrow$ BatchNorm $\rightarrow$ \\
3. FC($16 \, n_{\mathrm{out}}, \, n_{\mathrm{inference}}) \rightarrow$ Parameter means + Covariance matrix, \\

where $n_\mathrm{inference}$ is the number of values needed to be predicted by the model to construct both parameter means and covariance matrix:
\begin{equation}\label{eq:n_inference}
    n_{\mathrm{inference}} = n_{\mathrm{params}} + \frac{n_{\mathrm{params}}*(n_{\mathrm{params}} + 1)}{2}
\end{equation}

\section{Neural Network Architectures for Marginalization over Baryonic Effects}\label{appendix:Baryons_NN_Info}

\subsection{Encoder Network}
The encoder network consists of a sequence of fully-connected layers with varying numbers of hidden units (in this case, either 16 or 32). We apply batch normalization and ReLU activation function after each layer , except the output layer. The final output is the summaries $S$. We choose the dimensionality of the summaries to be equal to 32 ($dim(S) = 32$). 
\begin{equation}
    P(k) \rightarrow 16 \rightarrow 32 \rightarrow 32 \rightarrow 32 \rightarrow 32 \rightarrow 32 \rightarrow S
\end{equation}
\subsection{Inference Network}
The inference network is a simple fully-connected network with 2 hidden layers with $4\times dim(S)$. Similarly to the encoder network, we apply batch normalization and ReLU activation function after each layer, excluding the output layer. The final output is of dimensionality $n_{\mathrm{inference}}$ which is given by Eq. \ref{eq:n_inference}. 
\begin{equation}
    S \rightarrow 4\times dim(S) \rightarrow 4\times dim(S) \rightarrow n_{\mathrm{inference}}
\end{equation}

\section{Alternative self-supervised learning approaches: SimCLR}\label{appendix:SimCLR_vs_VICReg}
In this appendix we show results of data compression and parameter inference for the SimCLR method. Proposed by \cite{chen2020simple}, SimCLR is a contrastive learning method which stands for `A Simple Framework for Contrastive Learning of Visual Representations'. Similarly to VICReg and other self-supervised learning methods, SimCLR method consists of a pre-training step and a downstream task.

In the pre-training step, the encoder network samples a batch of $N$ inputs and uses two different views $X_i$ and $X_j$ corresponding to each input image $I$ in the batch. It treats the two views of the same image as a positive pair and consider all other examples in the batch as negative examples. The encoder network then learns the representations of the inputs by maximizing the similarity between the positive pair via a contrastive objective function which computed on the  embeddings $Z_i, Z_j$: 
\begin{equation}\label{eq:SimCLR_loss_pair}
    \ell_{i, j} = -\log \frac{\exp (\mathrm{sim} (Z_i, Z_j)/\tau)}{\sum_{k=1}^{2N}\mathbbm{1}_{[k \neq i]}\exp (\mathrm{sim} (Z_i, Z_j)/\tau)},
\end{equation}
where $\ell_{i, j}$ is the loss computed for each for a positive pair of examples, $\mathbbm{1}_{[k \neq i]}$ is an indicator function which is non-zero only if $k=i$, and $\tau$ is the so-called temperature parameter. In the following sections, we fix $\tau$ to 0.1.

The final loss function is computed by summing over all positive pairs in the batch:
\begin{equation}\label{eq:SimCLR_loss_batch}
    \mathcal{L} = \frac{1}{2N} \sum_{k=1}^{N}[\ell_{2k-1, 2k} + \ell_{2k, 2k-1}]
\end{equation}

While we outlined the motivation for using the VICReg method in Sec. \ref{sec:SSL}, it is interesting to check whether a contrastive learning method such as SimCLR would provide comparable results when applied to the mock cosmological datasets considered in this work. In this appendix we test the method on lognormal overdensity maps and CAMELS total matter density maps. We find that the results of parameter inference with SimCLR are comparable to the VICReg results which suggests that SimCLR could also offer an promising approach for data compression and inference in cosmology, although a further, more detailed experimentation with the method should be conducted.

\subsection{Lognormal Fields}
\subsubsection{SimCLR Data and Setup}
We use the dataset described in Sec. \ref{sec:LognormalFields_Data} and the encoder and inference network architectures outlined in Sec. \ref{sec:LognormalFields_VICReg}. 

We train the encoder network for 200 epochs in the \texttt{PyTorch} \citep{PytorchDLL} framework with \texttt{AdamW} \citep{kingma2014adam,hutter2019AdamW} optimizer with learning rate of $2\times 10^{-4}$, multiplied by a factor of 0.3 when the validation loss plateaus for 10 epochs. The downstream inference network is also trained for 200 epochs with \texttt{AdamW} optimizer, with the initial learning rate $5\times 10^{-4}$, reduced by a factor of 5 when the validation loss plateaus for 10 epochs.

\subsubsection{Results}
\begin{figure*}
\centering
 \includegraphics[width=\textwidth]{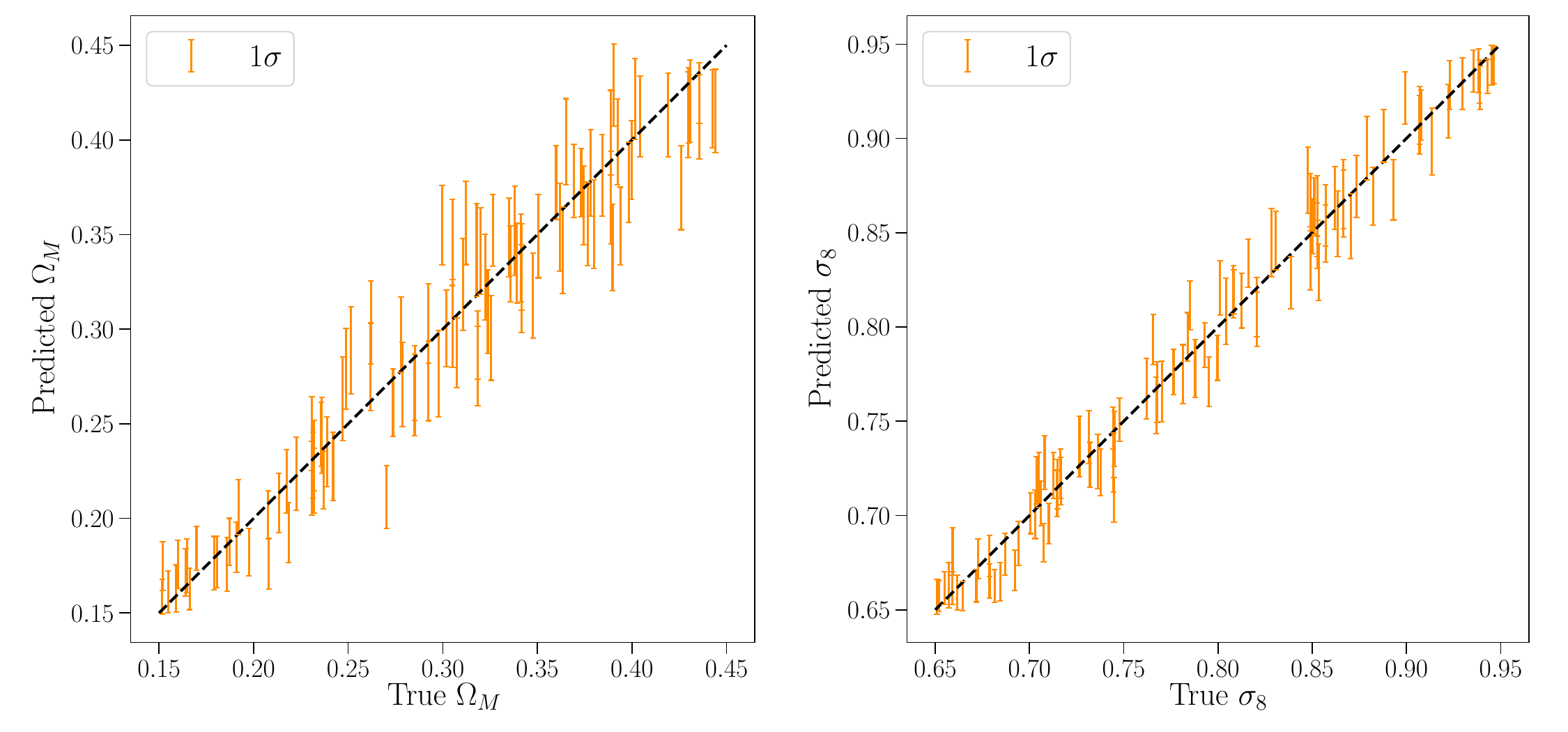}
 \caption{Scatter plot of predicted means and 1-$\sigma$ uncertainties of cosmological parameters $\Omega_M$ (left) and $\sigma_8$ (right) plotted against the true values of the parameters for a subset of a 100 maps from the test set. Predictions for the means and variances of the parameters were obtained by training a simple inference neural network on SimCLR summaries.}
 \label{fig:SimCLR_LN_fields_truth_vs_prediction}
\end{figure*}

We present the results of the parameter inference on the test dataset in Fig. \ref{fig:SimCLR_LN_fields_truth_vs_prediction} and compare the performance of the SimCLR and VICReg methods in Table \ref{tab:compare_VICReg_SimCLR}. 
We find that both methods are able to effectively compress the maps and accurately infer cosmological parameters from the summaries with similar errors. Once both the encoder and the inference networks are trained, we save the models with the lowest validation loss.

\begin{table}
 \centering
 \begin{tabular}{c c c c c}
  \hline
  \bf{Method} & \bf{Loss} & \bf{MSE} & \bf{MSE on $\Omega_M$} & \bf{MSE on $\sigma_8$} \\
   & & & (Relative error) & (Relative error) \\ \hline
  VICReg & -5.98 & 2.7$\times 10^{-4}$ & 3.6$\times 10^{-4}$  & 1.8$\times 10^{-4}$ \\
    & & & (5.2\%) & (1.3\%) \\ \hline
  SimCLR & -5.92  & 2.7$\times 10^{-4}$ & 3.7$\times 10^{-4}$ & 1.8$\times 10^{-4}$ \\
  & & & (5.2\%) & (1.3\%) \\
  \hline
 \end{tabular}
 \caption{Summary of the performance of the VICReg and SimCLR methods for inferring cosmological parameters $\Omega_M$ and $\sigma_8$ from lognormal overdensity maps. The performance of the two methods is evaluated on the test dataset described in Sec. \ref{sec:LognormalFields_Data}.}
 \label{tab:compare_VICReg_SimCLR}
\end{table}

\begin{figure}
 \includegraphics[width=\columnwidth]{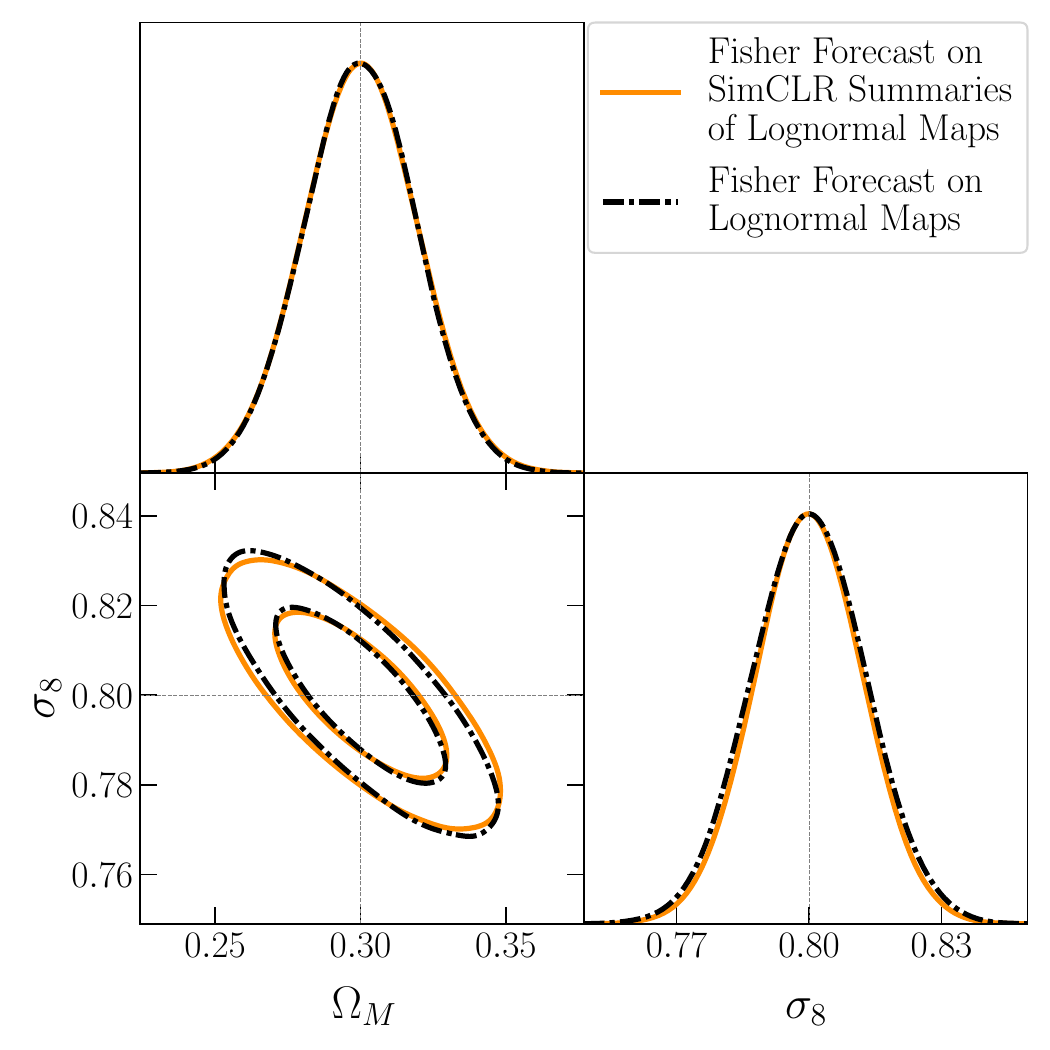}
 \caption{Constraints from Fisher forecast on the cosmological parameters $\Omega_M$ and $\sigma_8$ obtained from lognormal overdensity maps (black dash-dotted line) and from summaries constructed with VICReg (orange solid line). The results shown on the plot were obtained for a fiducial cosmology with $\Omega_M = 0.3$ and $\sigma_8 = 0.8$.}
 \label{fig:SimCLR_LN_fields_fisher_forecast}
\end{figure}

We also compare the Fisher information content of the lognormal fields and the SimCLR summaries, following the calculations in Sec. \ref{sec:LognormalFields_Fisher}.
We show the Fisher-informed constraints on cosmological parameters in Fig. \ref{fig:SimCLR_LN_fields_fisher_forecast} with fiducial values $\Omega_M=0.3$ and $\sigma_8=0.8$. We find the Fisher-informed contours from the lognormal fields and the SimCLR summaries to agree very well.

\subsection{CAMELS Total Matter Density Maps}
\subsubsection{SimCLR Data and Setup}
We next apply the SimCLR method to total matter density maps from the CAMELS project. The neural network architecture and datasets are same as in Sec. \ref{sec:CAMELS_data_vicreg}. 

We compare the results of the parameter inference with SimCLR to the VICReg results. The VICReg setup is the same as in Sec. \ref{sec:CAMELS_data_vicreg}, except that we use only one pair of views. We choose to use one pair of views for a one-to-one comparison, since the SimCLR is defined for a single positive pair of examples.

We train the SimCLR encoder network for 150 epochs with \texttt{AdamW} \citep{kingma2014adam,hutter2019AdamW} optimizer with initial learning rate of $1\times 10^{-3}$, multiplied by a factor of 0.3 when the validation loss plateaus for 10 epochs. The downstream inference network is trained for 200 epochs with \texttt{AdamW} optimizer, with the initial learning rate $1\times 10^{-3}$ and the same learning rate schedule as the encoder. We use the same setup to train the VICReg encoder and inference network, but decrease the initial learning to $5\times 10^{-4}$ when training the inference network.

\subsubsection{Results}
\begin{figure*}
\begin{subfigure}{\textwidth}
    \centering
    \includegraphics[width=\textwidth]{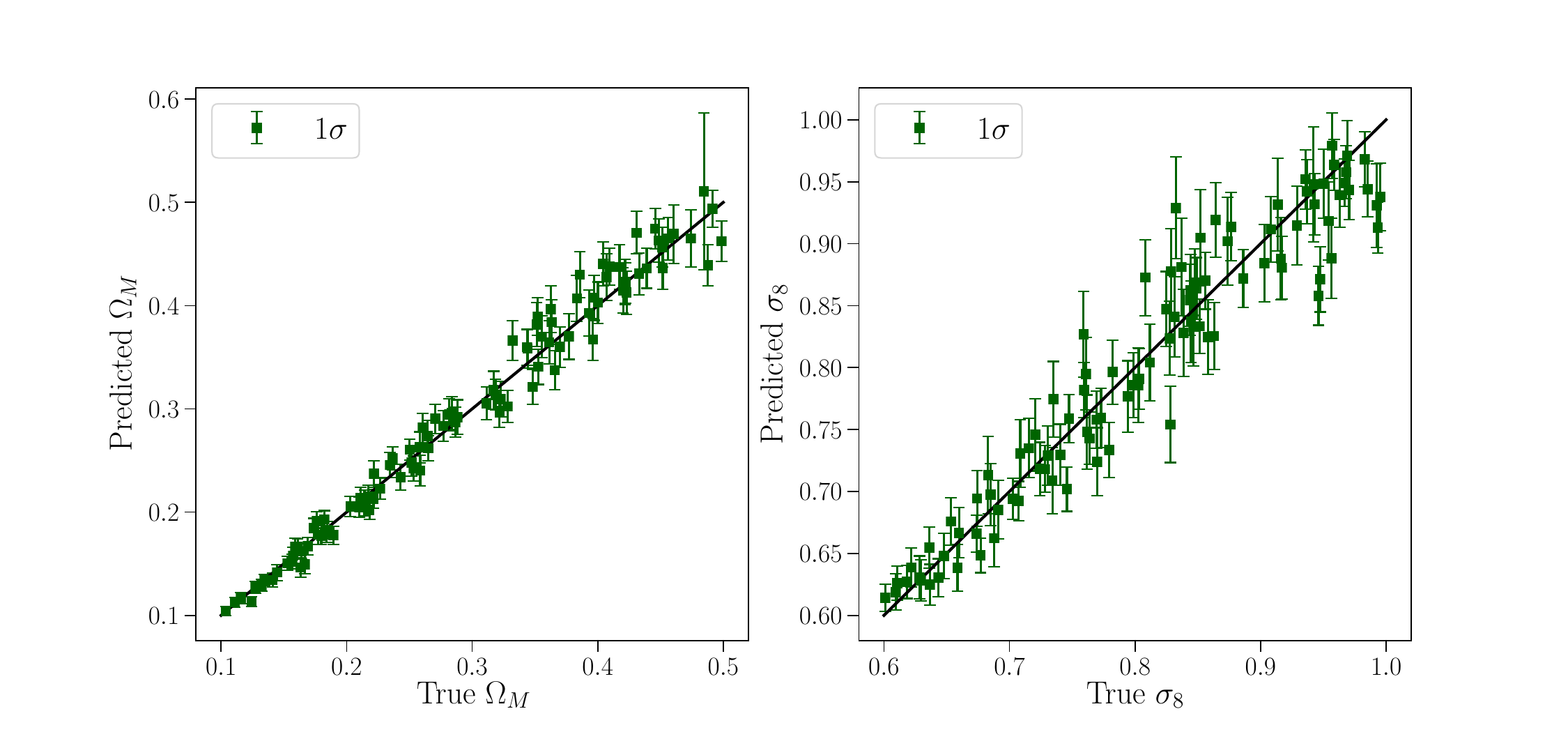}
    \caption{SimCLR: SIMBA.}
    \label{fig:SIMBA_fields_truth_vs_prediction_SimCLR}
\end{subfigure}
\begin{subfigure}{\textwidth}
    \centering
    \includegraphics[width=\textwidth]{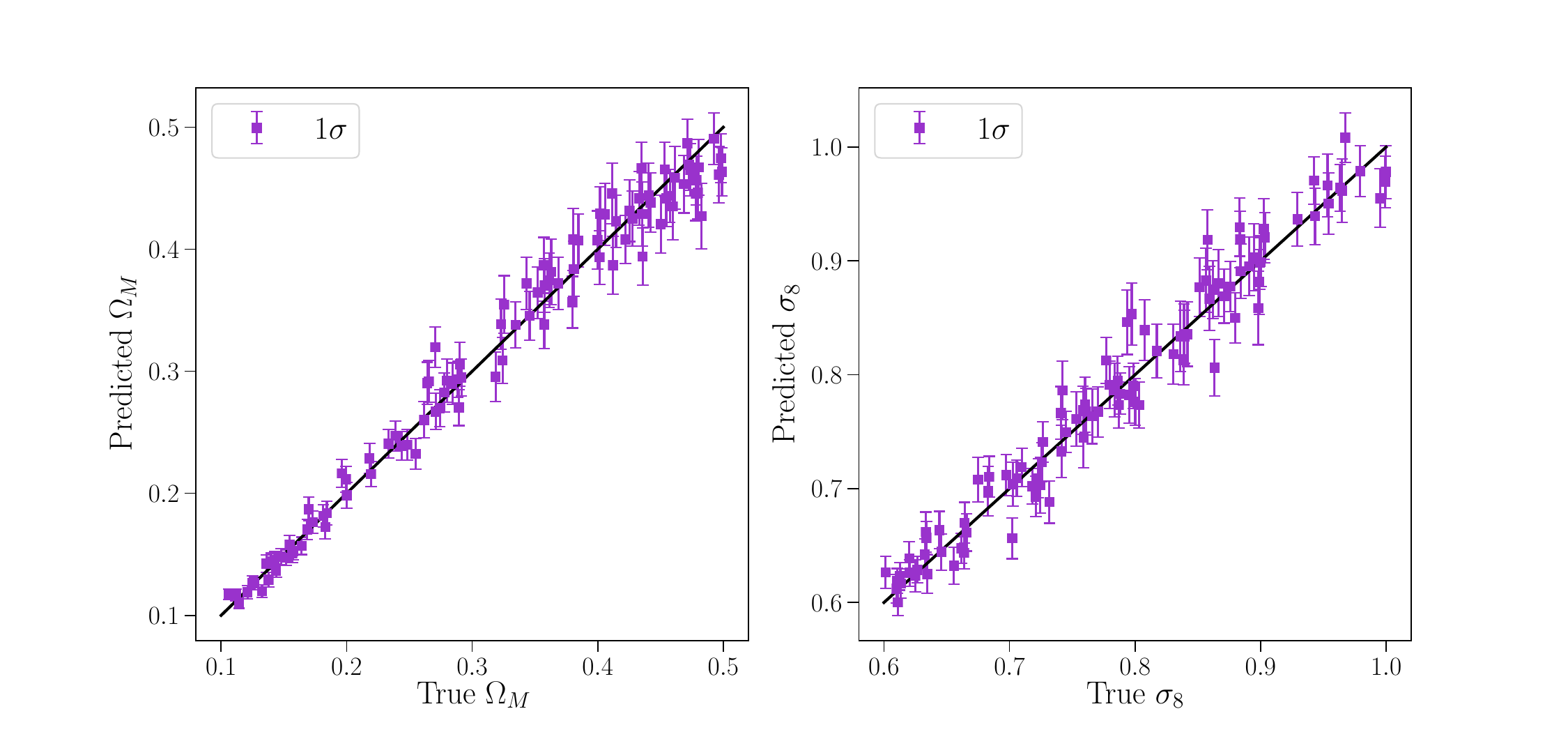}
    \caption{SimCLR: IllustrisTNG.}
    \label{fig:IllustrisTNG_fields_truth_vs_prediction_SimCLR}
\end{subfigure}       
\caption{Predicted means and 1-$\sigma$ uncertainties of cosmological parameters $\Omega_M$ and $\sigma_8$ compared to the true values of the parameters for total matter density maps from SIMBA and IllustrisTNG simulations. Predictions for the means and variances of the parameters were obtained by training simple inference neural network on SimCLR summaries.}
\label{fig:SIMBA_IllustrisTNG_fields_truth_vs_prediction_SimCLR}
\end{figure*}

\begin{figure*}
\begin{subfigure}{\textwidth}
    \centering
    \includegraphics[width=\textwidth]{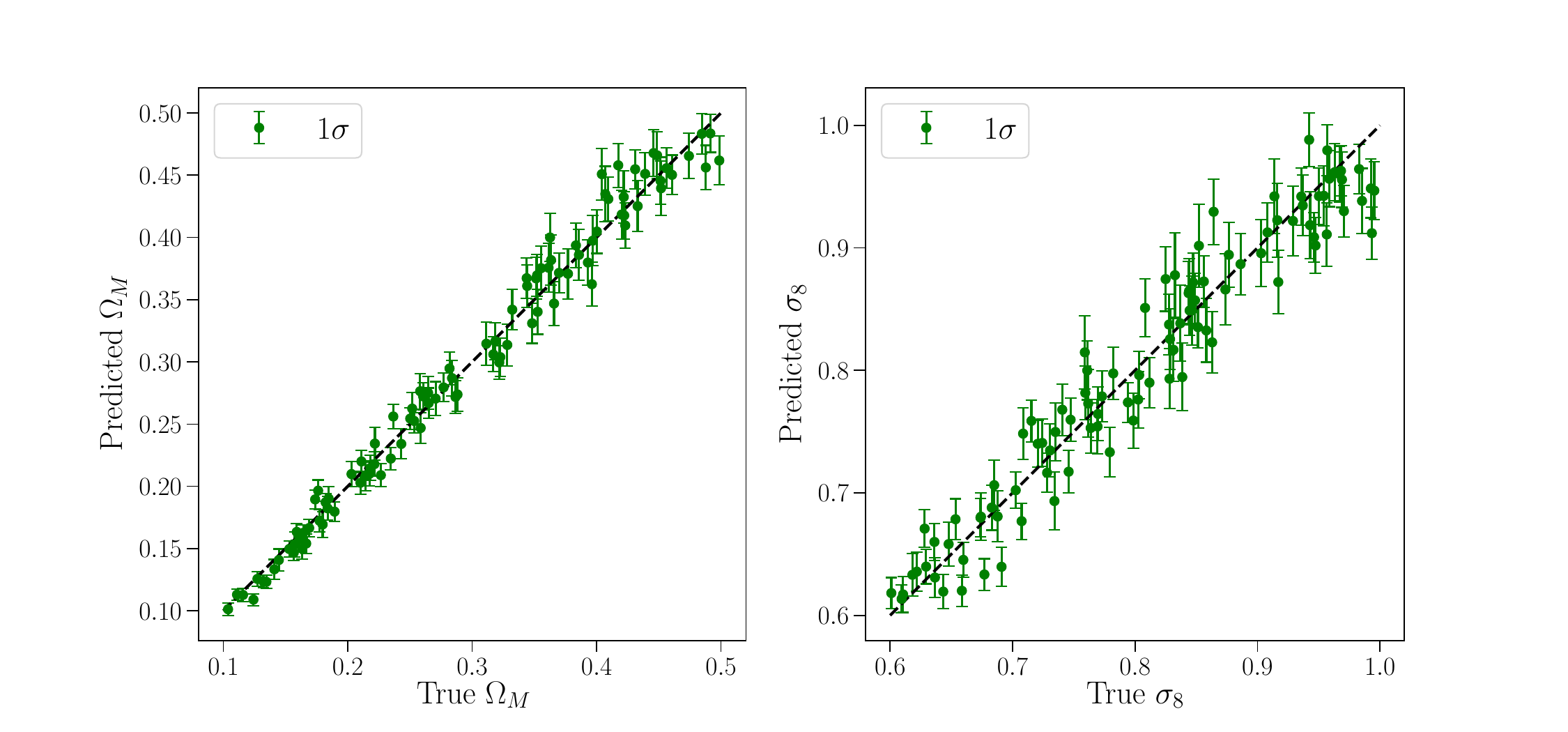}
    \caption{VICReg with 2 views: SIMBA.}
    \label{fig:SIMBA_fields_truth_vs_prediction_VICReg_2_views}
\end{subfigure}
\begin{subfigure}{\textwidth}
    \centering
    \includegraphics[width=\textwidth]{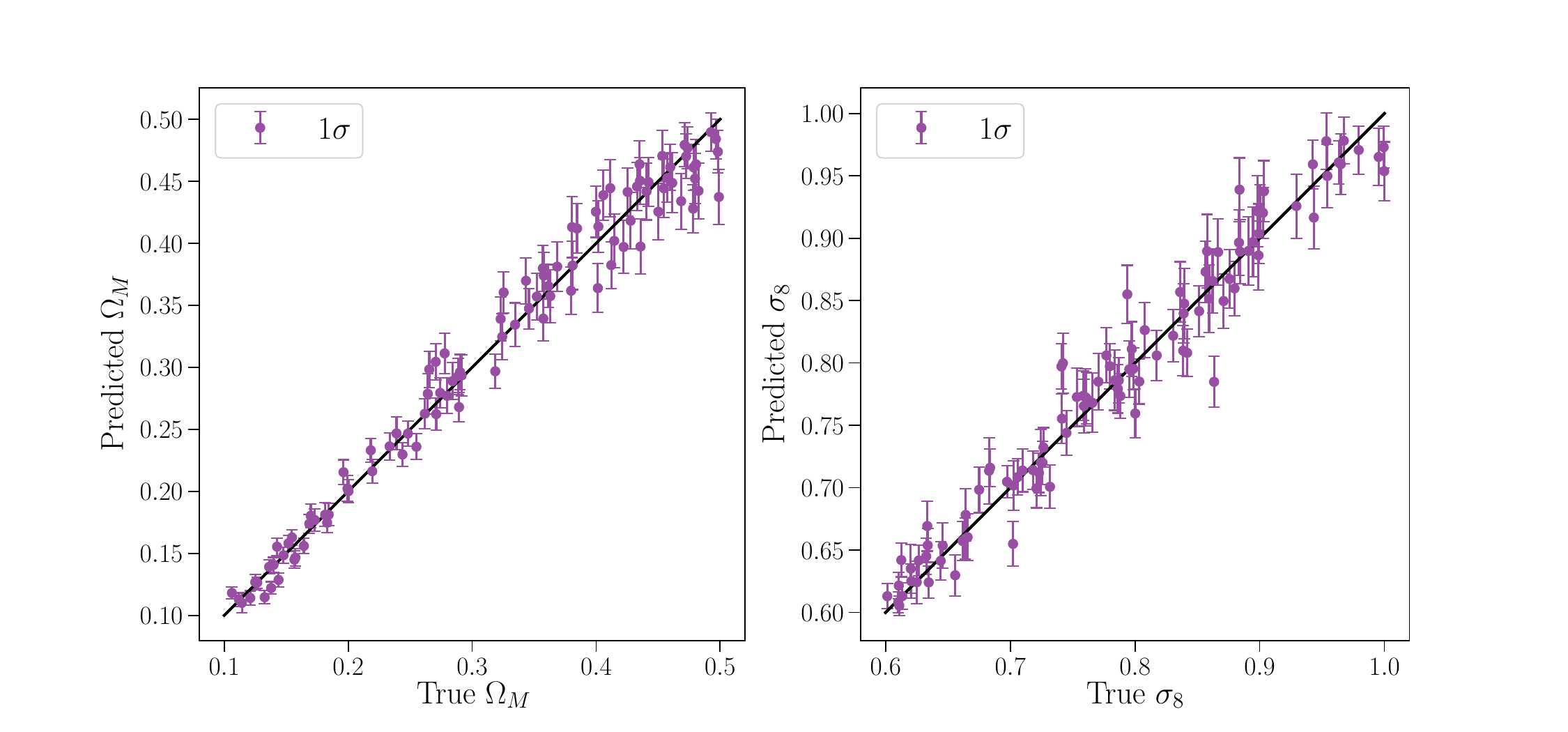}
    \caption{VICReg with 2 views: IllustrisTNG.}
    \label{fig:IllustrisTNG_fields_truth_vs_prediction_VICReg_2_views}
\end{subfigure}       
\caption{Predicted means and 1-$\sigma$ uncertainties of cosmological parameters $\Omega_M$ and $\sigma_8$ compared to the true values of the parameters for total matter density maps from SIMBA and IllustrisTNG simulations. Predictions for the means and variances of the parameters were obtained by training simple inference neural network on VICReg (2 views) summaries.}
\label{fig:SIMBA_IllustrisTNG_fields_truth_vs_prediction_VICReg_2_views}
\end{figure*}

We present our results for parameter inference with the two methods on the SIMBA and IllustrisTNG simulations suites in Fig. \ref{fig:SIMBA_IllustrisTNG_fields_truth_vs_prediction_SimCLR} (SimCLR) and \ref{fig:SIMBA_IllustrisTNG_fields_truth_vs_prediction_VICReg_2_views} (VICReg with 2 views). 
We summarize the losses and errors of SimCLR and VICReg methods on the test datasets in Tables \ref{tab:compare_VICReg_SimCLR_SIMBA} and \ref{tab:compare_VICReg_SimCLR_IllustrisTNG}. The results show that the two methods are comparable in terms of their performance and constraints on cosmological parameters.

\begin{table}
\begin{subtable}[h]{\columnwidth}
 \centering
 \begin{tabular}{c c c c c}
  \hline
  \bf{Method} & \bf{Loss} & \bf{MSE} & \bf{MSE on $\Omega_M$} & \bf{MSE on $\sigma_8$} \\
   & & & (Relative error) & (Relative error) \\ \hline
  VICReg & -3.22 & 4.62$\times 10^{-4}$ & 2.52$\times 10^{-4}$  & 6.73$\times 10^{-4}$ \\
  (2 views)  & & & (4.05\%) & (2.53\%) \\ \hline
  SimCLR & -3.24  & 5.14$\times 10^{-4}$ & 2.73$\times 10^{-4}$ & 7.55$\times 10^{-4}$ \\
  & & & (4.14\%) & (2.63\%) \\
  \hline
 \end{tabular}
 \caption{SIMBA.}
 \label{tab:compare_VICReg_SimCLR_SIMBA} 
\end{subtable}

\begin{subtable}[h]{\columnwidth}
\begin{tabular}{c c c c c}
  \hline
  \bf{Method} & \bf{Loss} & \bf{MSE} & \bf{MSE on $\Omega_M$} & \bf{MSE on $\sigma_8$} \\
   & & & (Relative error) & (Relative error) \\ \hline
  VICReg & -3.36 & 3.83$\times 10^{-4}$ & 3.12$\times 10^{-4}$  & 4.54$\times 10^{-4}$ \\
  (2 views)  & & & (4.17\%) & (2.07\%) \\ \hline
  SimCLR & -3.43  & 3.96 $\times 10^{-4}$ & 3.31$\times 10^{-4}$ & 4.60$\times 10^{-4}$ \\
  & & & (4.18\%) & (2.11\%) \\
  \hline
 \end{tabular}
 \caption{IllustrisTNG.}
 \label{tab:compare_VICReg_SimCLR_IllustrisTNG} 
 \end{subtable}
 \caption{Summary of the performance of the VICReg and SimCLR methods for inferring cosmological parameters $\Omega_M$ and $\sigma_8$ from SIMBA  and IllustrisTNG total matter density maps, evaluated on the respective test datasets, as described in Sec. \ref{sec:CAMELS_data_vicreg}.}
\end{table}

\section{Comparison to Theoretical Expectation (Fisher-Informed Constraints) with Normalizing Flows} \label{appendix:Fisher_forecast_examination}

The Fisher-informed constraints from the VICReg summaries of lognormal overdensity fields are computed under a set of assumptions about the Gaussianity of the likelihood of the summaries and of the posterior distribution of the parameters. In this Appendix, we confirm that our conclusions from Section \ref{sec:LognormalFields_Fisher} hold when we estimate the posterior distribution of the parameters from the VICReg summaries without making such assumptions.

We train a normalizing flow to estimate the distribution of the parameters, given a VICReg summary of a corresponding lognormal field. 
The normalizing flow consists of a stack of 8 masked autoregressive flows. 
We train the flow on VICReg summaries of the lognormal maps from the dataset described in Section \ref{sec:LognormalFields_Data}. 
We train the flow for 300 epochs using \texttt{AdamW} optimizer with learning rate of 5$\times 10^{-4}$ and cosine annealing schedule. 

We then compare the posteriors estimated by the normalizing flow to the constraints from Fisher forecast on the cosmological parameters $\Omega_M$ and $\sigma_8$ obtained from lognormal overdensity maps, following the calculation described in Section \ref{sec:LognormalFields_Fisher}.

We consider four lognormal overdensity maps for input parameters $\Vec{\theta}_o = \{\Omega_M = 0.3, \, \sigma_8=0.8\}$ with different initial random seeds. Since each map is a random realization of the fiducial cosmology, we expect the posteriors to be off-centered from the fiducial parameter values. In order to make a fair comparison with the Fisher-informed constraints for each map, we compute the Fisher matrix at the means of the parameter values inferred by the normalizing flow (as opposed to the fiducial cosmology values $\Vec{\theta}_o = \{\Omega_M = 0.3, \, \sigma_8=0.8\}$).

\begin{figure*}
\begin{subfigure}{0.4\textwidth}
    \centering
    \includegraphics[width=\textwidth]{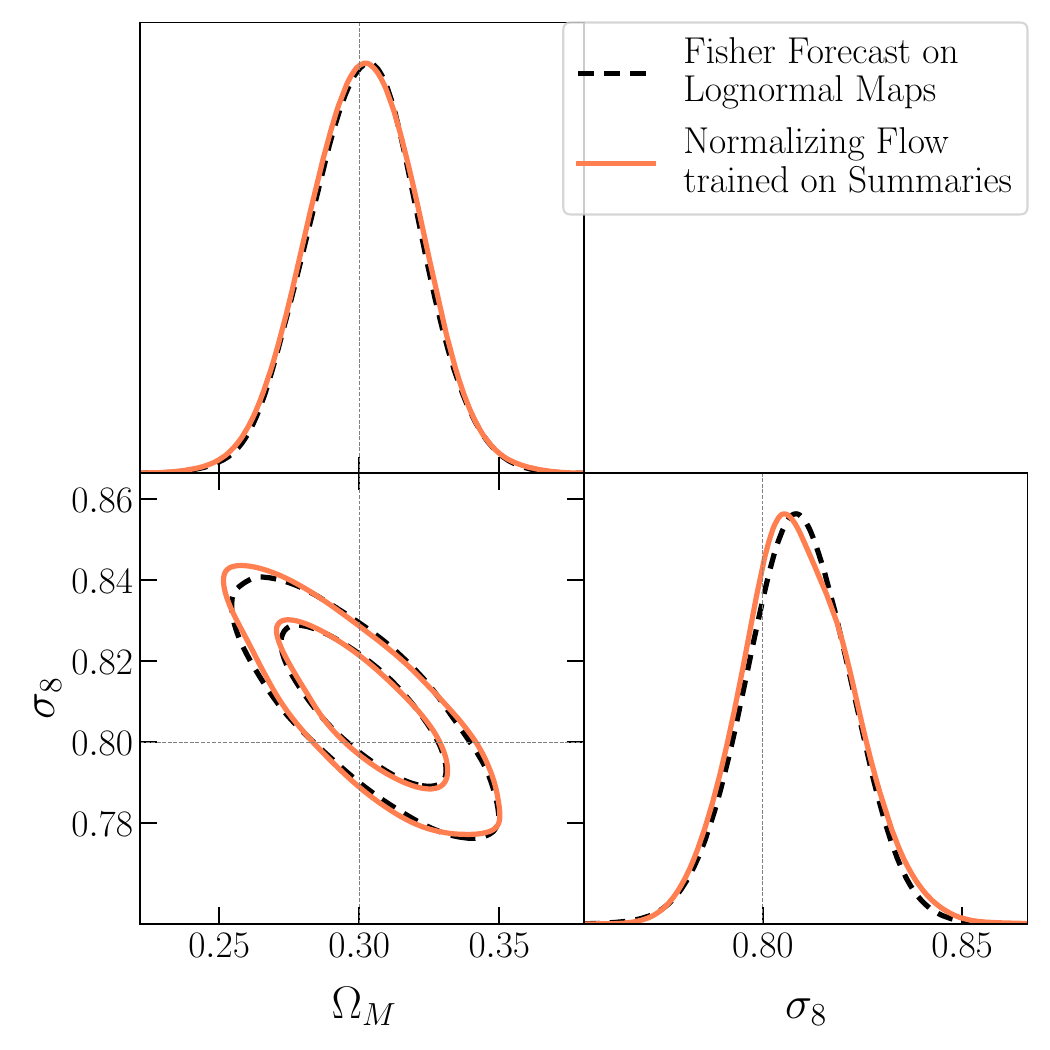}
\end{subfigure}
\begin{subfigure}{0.4\textwidth}
    \centering
    \includegraphics[width=\textwidth]{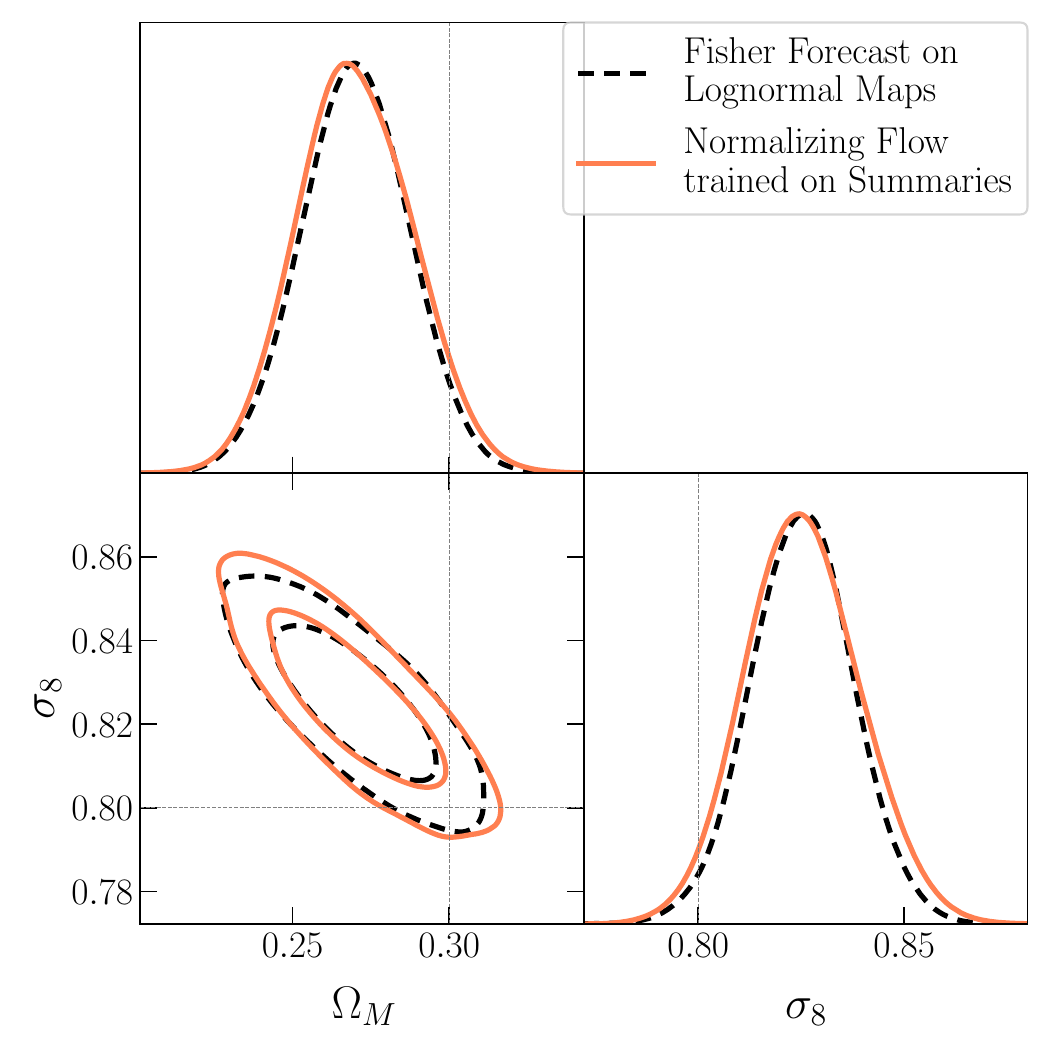}
\end{subfigure} 
\begin{subfigure}{0.4\textwidth}
    \centering
    \includegraphics[width=\textwidth]{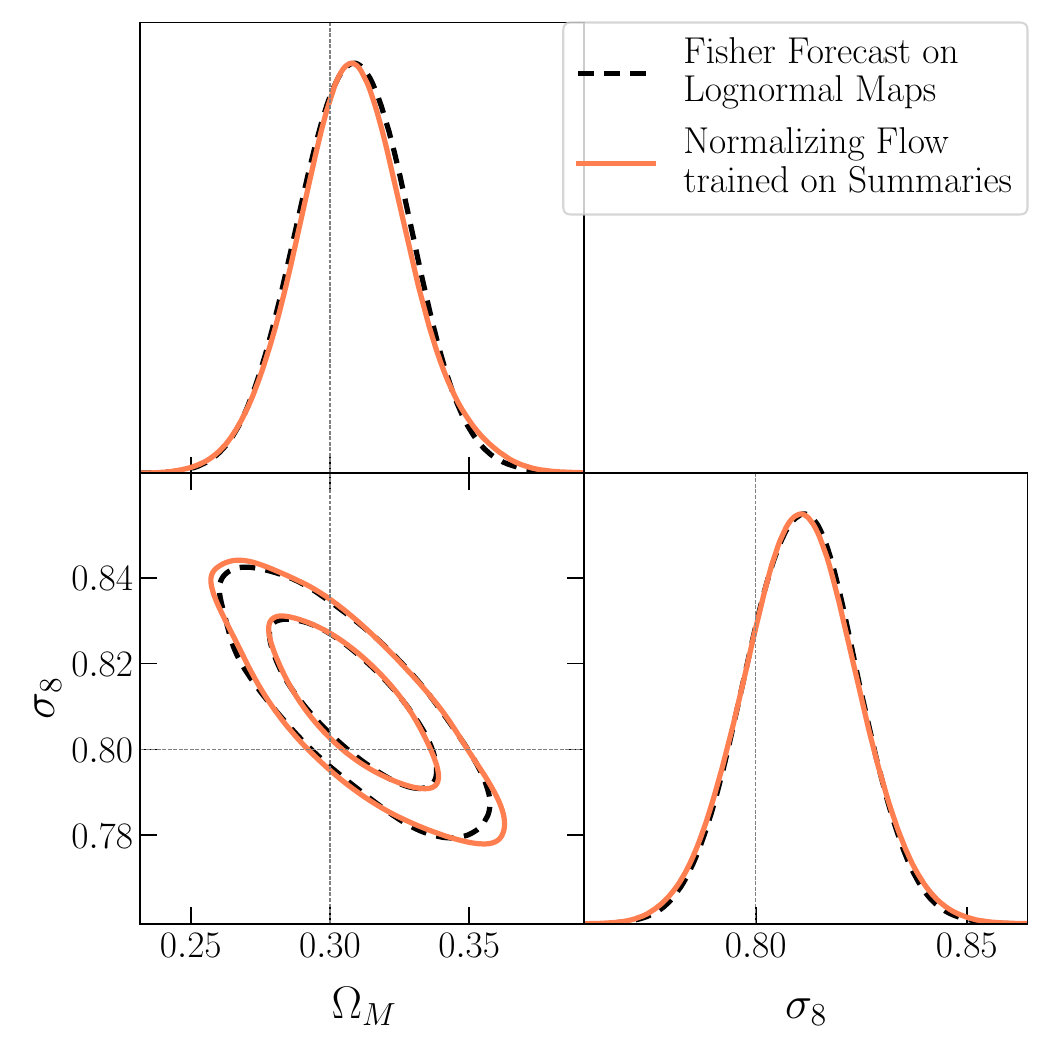}
\end{subfigure}
\begin{subfigure}{0.4\textwidth}
    \centering
    \includegraphics[width=\textwidth]{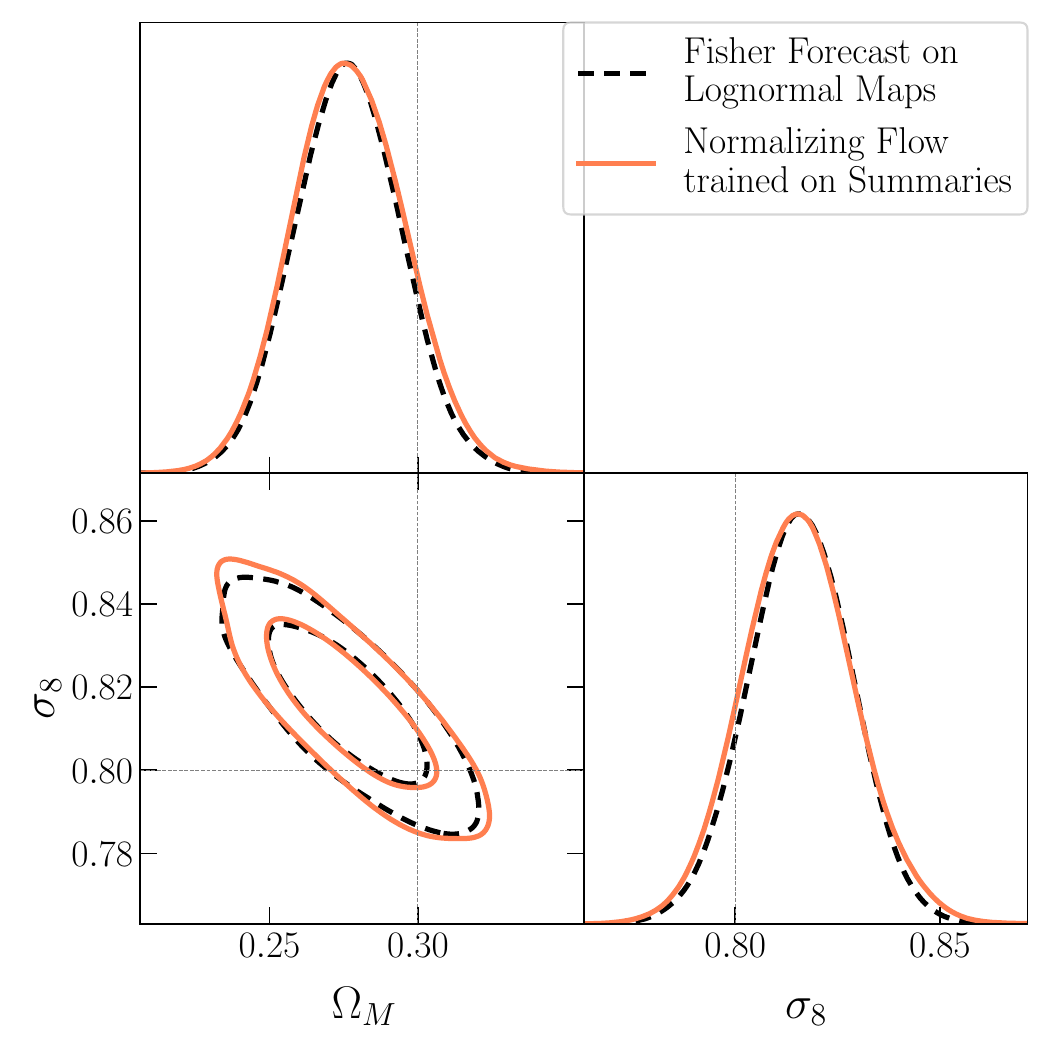}
\end{subfigure}
\caption{Constraints from Fisher forecast on the cosmological parameters $\Omega_M$ and $\sigma_8$ obtained from lognormal overdensity maps (black dash-dotted line) and inferred posterior distribution of parameters $\Omega_M$, $\sigma_8$ from a normalizing flow trained on the VICReg summaries (orange solid line). 
The results shown on the plot were obtained for four different random realizations of the lognormal fields with fiducial parameter values of $\Omega_M = 0.3$ and $\sigma_8 = 0.8$ (marked with a dashed line).}
\label{fig:Fisher_nflow}
\end{figure*}

We show the comparison for the four maps on Figure \ref{fig:Fisher_nflow}. As can be seen from the plot, posteriors inferred with the normalizing flow closely follow constraints from the Fisher forecasts, consistent with our conclusion in Section \ref{sec:LognormalFields_Fisher}.

\section{Posterior Calibration Tests} \label{appendix:posterior_calib}
We examine whether the SNPE-inferred posterior and the posterior obtained using the inference network from Section \ref{sec:application_to_SBI} are well-calibrated. 
We use the `simulation-based coverage calibration' (SBCC) procedure \cite{sbi_deistler2022truncated} as our coverage diagnostic. 
For a given parameter-data pair $\{\boldsymbol{\theta}, \boldsymbol{\mathrm{x}}\}$, SBCC estimates the coverage of the approximate posterior as follows: 
\begin{equation}
\mathrm{Coverage}=\int q_\phi(\boldsymbol{\theta} | \mathbf{x}^*) \mathcal{H} (\boldsymbol{\theta}) d \boldsymbol{\theta} \, ,
\end{equation}
where
$\mathcal{H}(\boldsymbol{\theta}) =
\begin{cases}
    1, & \text{if  } q_\phi(\boldsymbol{\theta}^* | \mathbf{x}^*) \geq q_\phi(\boldsymbol{\theta} | \mathbf{x}^*)\\
    0,             & \text{otherwise.}
\end{cases}$

The expected coverage is then computed as an average coverage across for multiple pairs of across multiple parameter-data pairs $\{\boldsymbol{\theta}^*, \boldsymbol{\mathrm{x}}^*\}$.
If the posterior is well-calibrated posterior, the expected coverage should match the confidence level for all confidence levels $(1-\alpha) \, \in [0, 1]$. 
The algorithm to compute SBCC is described in detail in the Appendix of \cite{sbi_deistler2022truncated}.

Since the posterior we obtain with simulation-based inference is not amortized, but targeted to a specific data point or observation (namely, a VICReg summary of a specific lognormal map), we do not expect the posterior to be accurate for all parameter-data pairs drawn from the prior. 
Rather, we would like the posterior to be well-calibrated for other observations that share the same fiducial cosmology as the target observation. 
In this case, fiducial cosmology corresponds to $\boldsymbol{\theta_o} = \{\Omega_M = 0.3, \, \sigma_8=0.8\}$.
Therefore, in order to compute the calibration plots, we use an emulator to sample 1000 different VICReg summaries at the fiducial parameter values and evaluate the coverage diagnostic on these samples.

\begin{figure*}
\begin{subfigure}{0.45\textwidth}
    \centering
    \includegraphics[width=\textwidth]{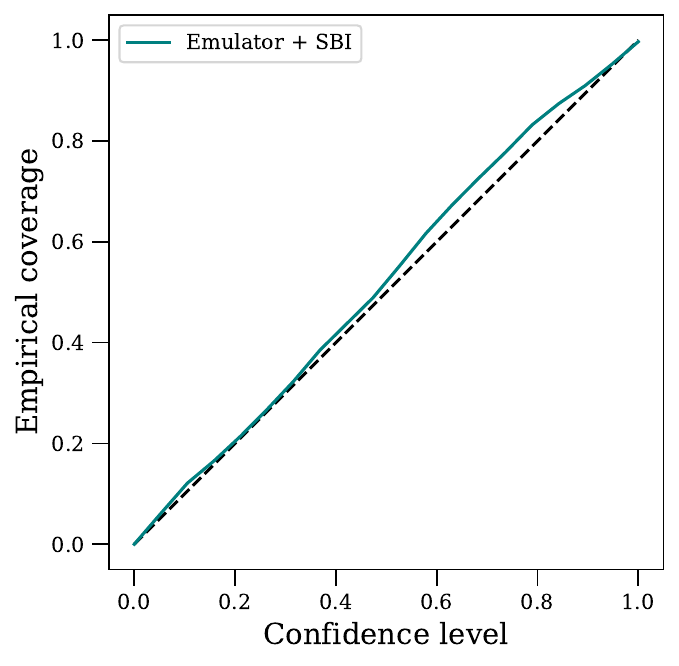}
    \caption{SNPE}
    \label{fig:emulator_prev_coverage}
\end{subfigure}
\begin{subfigure}{0.45\textwidth}
    \centering
    \includegraphics[width=\textwidth]{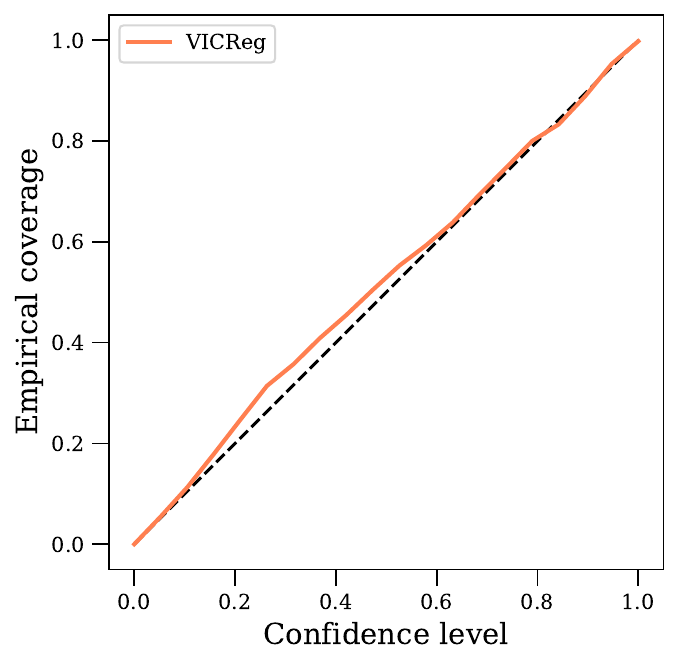}
    \caption{VICReg}
    \label{fig:emulator_VICReg_coverage}
\end{subfigure} 
\caption{Expected coverage versus confidence levels computed for the SNPE-obtained posteriors (left) and the VICReg posterior (right) from Section \ref{sec:application_to_SBI}.}
\label{fig:emulator_coverage}
\end{figure*}

We plot the expected coverage for a range of confidence levels $(1-\alpha)$ for the two posteriors on Figure \ref{fig:emulator_coverage}. As can be seen from the plot, the both posteriors lie close the dashed line which corresponds to a well-calibrated posterior.

\bsp
\label{lastpage}
\end{document}